Research paper

# Mixed convection and entropy generation analysis of carbon nanotube-water nanofluid in a square cavity with cylinders and flow deflectors

Hashnayne Ahmed [a,b,*], Shashanka Biswas [a], Farzana Akter Tina [a]

[a] Department of Mathematics, Faculty of Science and Engineering, University of Barishal, Barishal, 8200, Bangladesh
[b] Department of Mechanical and Aerospace Engineering, University of Florida, Gainesville, 32611, FL, USA

ABSTRACT

This study explores the mixed convection of carbon nanotube (CNT)-water nanofluid within a square cavity containing heated cylinders under the influence of a magnetic field, focusing on three geometric configurations: a single heated cylinder, two heated cylinders, and two heated cylinders with a flow deflector. The impact of various parameters, including Reynolds number ($Re$), Richardson number ($Ri$), Hartmann number ($Ha$), wavy wall peaks ($n$), nanoparticle volume fraction ($\phi$), Hartmann angle ($\gamma$), rotational speed ($\omega$), and inclination angle ($\alpha$), on thermal and fluid dynamic behaviors is analyzed. MWCNT nanofluids exhibit up to a 19.1% increase in $Nu_{ave}$ compared to SWCNT nanofluids, confirming their superior heat transfer performance. Adding a second heated cylinder increases $Nu_{ave}$ by approximately 71.7% compared to a single-cylinder configuration, while the inclusion of a flow deflector modifies vortex structures, further enhancing convective transport. Increasing wavy wall peaks ($n$) enhances heat transfer by intensifying vortex formation and disrupting thermal boundary layers, leading to a more uniform temperature distribution. SWCNT nanofluids exhibit Bejan numbers up to 58.7% higher than MWCNT nanofluids, indicating greater thermal irreversibility. These findings provide valuable insights for optimizing thermal management systems in engineering applications, highlighting the importance of selecting appropriate nanofluids, geometric configurations, and magnetic field parameters to achieve optimal thermal performance and fluid stability.

## 1. Introduction

The study of convective heat transfer and fluid flow in enclosed spaces is crucial due to its broad range of engineering applications. The rapid advancement of electronic devices, solar cells, heating elements, heat exchangers, climate control, ventilation systems, electronics cooling, drying equipment, thermal energy storage systems, fuel cell technology, storage facilities, geothermal installations, nuclear and chemical reactors, food processing sectors, and lubrication technologies highlights the need for efficient methods to enhance heat transfer performance. To achieve this, the use of mixed nanoparticles is essential [1]. Nanostructures also play a crucial role in various engineering, technological, and natural processes applications [2–8]. Notably, these applications include cooling systems used in sectors like energy production, transportation, and semiconductors. Research has demonstrated that nanofluids significantly enhance heat transfer [9–13]. In recent decades, significant attention has been directed toward nanofluids, MEMS, and the structural, morphological, and optical characteristics of nanostructures [14–17]. Researchers have been particularly interested in these areas because free and forced convection heat transfer from nanofluids like CNT-water has numerous industrial applications, especially in cooling systems [18].

Enhancing the heat transfer capabilities of nanofluids has been a focal point in recent research. For example, Maatki investigated natural double diffusive convection in triangular solar panels utilizing CNT-water nanofluids and found that shorter CNT lengths led to lower average Nusselt numbers at the absorber plate [19]. Similarly, Omri et al. highlighted significant performance improvements in three-dimensional heat exchangers with triangular fins when CNT nanofluids were incorporated [20]. Hossain et al. examined MHD mixed convection of kerosene oil-based CNT nanofluids in square lid-driven cavities under unsteady radiative heat flow, discovering that increasing particle concentration enhances both the fluid's thermophysical properties and its heat transfer

* Corresponding author at: Department of Mathematics, Faculty of Science and Engineering, University of Barishal, Barishal, 8200, Bangladesh.
*E-mail addresses:* hashnayneahmed17@gmail.com (H. Ahmed), shashankabiswasbu@gmail.com (S. Biswas), farzanatina17@gmail.com (F.A. Tina).

**Nomenclature**

| | | |
|---|---|---|
| $a$ | amplitude of the wavy wall | |
| $B_0$ | magnetic field strength | $T$ |
| $Be$ | Bejan number | |
| $C_p$ | specific heat at constant pressure | $J kg^{-1} K^{-1}$ |
| $g$ | gravitational acceleration | $ms^{-2}$ |
| $Gr$ | Grashof number | $Re^2 \cdot Ri$ |
| $h$ | heat transfer coefficient | $W m^{-2} K^{-1}$ |
| $Ha$ | Hartmann number | |
| $k$ | thermal conductivity | $W m^{-1} K^{-1}$ |
| $L$ | length and height of the cavity | $m$ |
| $n$ | number of wavy wall peaks | |
| $Nu$ | Nusselt number | $hx/k$ |
| $P$ | dimensionless pressure | $p/\rho_{nf} U_0^2$ |
| $p$ | pressure | $Pa$ |
| $Pr$ | Prandtl number | $\nu_f/\alpha_f$ |
| $R$ | volume fraction of the nanoparticles | |
| $Ra$ | Rayleigh number | $g\beta_f L^3 \Delta T/\nu_f \alpha_f$ |
| $Re$ | Reynolds number | $U_0 L/\nu_f$ |
| $Ri$ | Richardson number | $Ra/Pr \cdot Re^2$ |
| $S_h$ | local entropy generation rate due to heat transfer | |
| $S_j$ | local entropy generation rate due to Joule heating | |
| $S_t$ | total entropy generation | |
| $S_v$ | local entropy generation rate due to fluid friction | |
| $T$ | temperature | $K$ |
| $U, V$ | dimensionless velocity components | |
| $u, v$ | velocity components in Cartesian coordinates | $ms^{-1}$ |
| $X, Y$ | dimensionless coordinates | |
| $x, y$ | Cartesian coordinates | $m$ |

*Greek Symbols*

| | | |
|---|---|---|
| $\alpha$ | thermal diffusivity | $m^2 s^{-1}$ |
| $\beta$ | thermal expansion coefficient | $K^{-1}$ |
| $\gamma$ | Hartmann angle | |
| $\mu$ | dynamic viscosity | $kg\, s/m$ |
| $\nu$ | kinematic viscosity | $\mu/\rho$ |
| $\omega$ | cylinder rotating speed | |
| $\phi$ | concentration ratio of the nanoparticles | |
| $\rho$ | density | $kg/m^3$ |
| $\sigma$ | electrical conductivity | |
| $\theta$ | dimensionless temperature | $(T - T_c)/(T_h - T_c)$ |

*Subscripts*

| | |
|---|---|
| $ave$ | average |
| $c$ | cold |
| $f$ | base fluid |
| $h$ | hot |
| $l$ | local |
| $nf$ | nanofluid |
| $s$ | solid particle |
| $t$ | total |
| $w$ | wall |

*Abbreviations*

| | |
|---|---|
| CNT | Carbon Nanotube |
| GFEM | Galerkin Finite Element Method |
| MWCNT | Multi-Wall Carbon Nanotube |
| SWCNT | Single-Wall Carbon Nanotube |

rate [21]. Ganesh et al. also observed that alumina-water nanofluids in square cavities with triangular barriers outperformed those with circular or square barriers in terms of heat transfer rates [22]. Hybrid nanofluids have shown significant potential in enhancing heat transfer. Nabi et al. employed hybrid nanofluids with turbulence-inducing features in flat plate solar collectors, achieving heat transfer coefficient increases of 8% and 4.1% for SWCNT-CuO water and MWCNT-CuO water, respectively, compared to pure water at a Reynolds number of $10^4$ [23]. Furthermore, hybrid nanofluids at lower concentrations improved the thermal performance of the collectors. Behzadnia et al. found that spherical alumina nanoparticles were most efficient for supercritical water reactor cooling models, providing the highest Nusselt numbers and optimal flow behavior [24].

Advancements in numerical methods have led to a deeper understanding of nanofluid behavior. Xia et al. used a two-phase mixture approach to model mixed convection of alumina-water nanofluid in a T-shaped lid-driven cavity, showing that higher Richardson numbers reduced heat transfer by diminishing secondary flows through the sliding lid. Additionally, increasing the aspect ratio of the cavity lowered heat transfer, whereas a higher nanoparticle volume fraction enhanced it [25]. Tayebi et al. studied micropolar alumina-water nanofluid in an inclined I-shaped enclosure with two heated cylinders, focusing on free convection and entropy generation. They found that dispersing the nanomaterial significantly boosted the thermal free exchange rate [26]. Tang et al. investigated a two-layer micro-sized heat sink with sinusoidal cavities, discovering that graphene nanoplatelet sodium dodecylbenzene sulfonate-water nanofluid increased the convective heat transfer coefficient at higher Reynolds numbers [27]. Rahmati et al. used the lattice Boltzmann technique to simulate mixed convection in a double lid-driven cavity, finding that reducing the Richardson number increased Nusselt numbers and heat transfer for all thermal conditions [28]. Hasan et al. discovered that higher Richardson numbers enhanced heat transfer rates in a square cavity filled with Cu-water nanofluid [29].

Research into various geometrical configurations has yielded valuable insights for optimizing heat transfer. Khan et al. investigated natural convection in a hexagonal enclosure with a heated cylinder containing CuO-water, finding that increasing the Darcy number improved heat transfer rates and velocity distributions [30]. Kolsi et al. evaluated hybrid nanofluids in bifurcating channels with double rotating cylinders and partially porous layers, noting that Nusselt numbers and pressure coefficients improved with higher solid volume percentages and rotational speeds [31]. Arif et al. showed that using a ternary hybrid nanofluid with various shapes of nanoparticles enhanced heat transfer rates by 33.67% in radiator applications [32]. Alhajaj et al. compared different setups (porous block, porous straight channel, and porous wavy channels) using various nanofluids. They found that each nanofluid's performance varied depending on the design, with 0.5% volume $Al_2O_3$-water offering the highest efficiency index in terms of Nusselt number and pressure drop [33]. Chougule et al. investigated heat transfer enhancement in a circular tube with CNT-water nanofluid and wire coil inserts, showing increased heat transfer rates with minor friction factor increases [34].

Studies on natural convection have highlighted significant findings regarding nanofluid behavior. Islam et al. examined natural convection in a triangular enclosure with a sinusoidal heat source, observing increased thermal and fluid activity with higher Rayleigh numbers [35]. Hirpho and Wubshet found that higher Rayleigh numbers shifted circulation centers towards the top wall in a trapezoidal enclosure with a heated inner circular cylinder filled with Casson nanofluid [36]. Similarly, Sameh et al. showed that increasing the nonlinear Boussinesq parameter in an inclined open arc-shaped channel filled with isotropic porous media enhanced the average Nusselt number [37]. Zarei et al. studied heat transport in a square cavity filled with nanofluid and surrounded by sinusoidal wavy walls. They concluded that the volume fraction was the most critical parameter affecting heat transfer, with the average Nusselt number increasing significantly as the volume frac-tion increased. Conversely, the wall wavelength had a minimal impact on heat transfer [38].

Mixed convection studies have demonstrated various methods for improving heat transfer. Refiei et al. improved the performance of the solar organic Rankine cycle using MWCNT-oil nanofluid, showing that the hemispherical cavity receiver with nanofluid achieved the highest system efficiency [39]. Uddin et al. used a non-uniform dynamic model to study natural convection of CuO-water nanofluid in a square domain with wavy vertical surfaces, finding significant increases in heat transfer with higher Rayleigh numbers and nanoparticle volume fractions [40]. Waini et al. studied mixed convection over an exponentially stretching vertical surface with hybrid nanofluid, finding that increased nanoparticle volume percentages for copper and alumina reduced the heat transfer rate [41]. Zhang et al. examined mixed convection and entropy generation in a semi-elliptic lid-driven cavity filled with Ag-water nanofluid, showing that increasing Richardson and Grashof numbers improved velocity and temperature distribution [42]. Raizah et al. found that increasing the Darcy number enhanced temperature gradients and average Nusselt numbers in a V-shaped cavity filled with nanofluid and heterogeneous porous media [43,44]. Eshagh et al. demonstrated that higher buoyancy ratios and lower Lewis numbers enhanced Nusselt and Sherwood numbers in an H-shaped enclosure with hybrid Cu–$Al_2O_3$–$H_2O$ nanofluid. Alsabery et al. studied free convection in a 2D wavy-walled container heated from below, showing that higher volume fractions optimized heat transfer rates and that temperature differences caused significant concentration changes [45,46].

Recent studies have focused on entropy generation and Bejan number calculations in various configurations. Bijan et al. [47] analyzed entropy generation in a cavity with a heated circular cylinder, showing that increasing the Nusselt and Bejan numbers enhances heat transfer and system efficiency. Iftikhar et al. studied MHD mixed convection flow of a non-Newtonian fluid in a square cavity, observing increases in fluid velocity, energy transfer rate, and entropy generation with higher viscosity. Conversely, a higher Hartmann number reduced velocity and temperature but raised entropy generation, with specific bi-viscosity, Prandtl, and Grashof numbers yielding minimal entropy generation [48]. Ozgun et al. [49] examined $Al_2O_3$-water nanofluid in a lid-driven square cavity, finding that increasing the nanoparticle volume fraction and lowering the Richardson number enhanced the Nusselt number and entropy generation. Studies have also explored the role of magnetic fields in controlling entropy generation and heat transfer, with Fayz-Al-Asad et al. [50] demonstrating that fin size significantly alters flow mixing and thermal performance in magneto-convection systems, while Iftikhar et al. [48] observed that higher Hartmann numbers suppress convective motion but increase entropy generation. Beyond single-phase nanofluids, hybrid nanofluids have been investigated for their improved heat transfer capabilities and entropy control, as shown by Mebarek-Oudina et al. [51], who examined MgO-$Al_2O_3$/water hybrid nanofluids in a porous elliptical cavity under a magnetic field. These studies collectively highlight how the combined effects of nanoparticle composition, geometric modifications, and magnetic field interactions influence heat transfer and entropy generation, aligning with the objectives of the present work.

In this study, the mixed convection and entropy generation analysis of CNT-water nanofluid in a square cavity with heated cylinders, under the influence of a magnetic field, is investigated. Three geometric configurations are analyzed: a single heated cylinder, two heated cylinders, and two heated cylinders with a flow deflector. The research evaluates the impact of various parameters, including Reynolds number ($Re$), Richardson number ($Ri$), Hartmann number ($Ha$), wavy wall peaks ($n$), nanoparticle volume fraction ($\phi$), Hartmann angle ($\gamma$), rotational speed ($\omega$), and inclination angle ($\alpha$), on thermal and fluid dynamic behaviors. The novelty of this study lies in its systematic evaluation of the combined effects of nanofluid properties, magnetic field influence, and geometric modifications on mixed convection and entropy generation. Unlike prior research, which primarily focuses on either heat transfer enhancement or entropy minimization separately, this study presents a comprehensive framework integrating both. Additionally, a more detailed assessment of the irreversibility factor ($\Theta$) is provided, demonstrating its role in quantifying energy dissipation at varying Reynolds numbers. Key findings reveal that MWCNT nanofluids exhibit superior heat transfer performance, as evidenced by higher Nusselt numbers, while SWCNT nanofluids show higher Bejan numbers, indicating greater thermal entropy generation. The study also highlights the significant role of magnetic fields in stabilizing flow and enhancing thermal performance, and the critical influence of geometric modifications such as additional cylinders and flow deflectors on optimizing fluid mixing and heat transfer efficiency. The results provide valuable insights into designing next-generation nanofluid-based thermal management systems applicable in aerospace, electronics cooling, and industrial heat exchangers. Practical relevance is further supported by emerging real-world use cases, including CNT-based thermal interface materials in spacecraft and high-performance computing systems [52,53], and commercial nanofluid products adopted in HVAC systems [54]. By connecting simulation insights with applied technologies, this work contributes a foundational understanding for engineering high-efficiency, nanofluid-enhanced thermal systems.

The remaining sections of this article follow a structured organization. In § 2, the physical problem is explained, accompanied by the formulation of relevant mathematics and the consideration of parameters essential for entropy calculation. § 3 delineates the numerical techniques utilized for problem resolution, coupled with the necessary validation process crucial for the progression of this project. Finally, § 4 showcases the simulation results, highlighting the characteristics of the two types of nanofluids and elucidating the influence of flow stabilizers within the domain.

## 2. Problem description

### 2.1. Geometry setup

As depicted in Fig. 1, the geometry of the problem consists of a two-dimensional square enclosure with equal length and height, denoted as $L$. Three distinct geometric configurations within a square cavity containing CNT-water nanofluid under the influence of a magnetic field are investigated. Each configuration varies by the number and placement of heated cylinders and the introduction of a flow deflector. Case 1 features a single heated cylinder positioned at the cavity's center, creating a baseline for thermal and fluid dynamic behavior. Case 2 introduces a second heated cylinder, leading to stronger thermal coupling and intensified convective interactions, enhancing heat transfer efficiency. Case 3 further enhances the system's complexity by incorporating a circular, insulated flow deflector that rotates with a unit speed, modifying vortex structures, reducing stagnant regions, and enhancing thermal transport.

The non-dimensional boundary conditions applied to the cavity, heated cylinder walls, flow deflector walls, and supportive insulator walls are summarized in Table 1. For the sake of efficient heat transfer coefficient analysis, two types of carbon nanotube (CNT)-water nanofluids are considered (SWCNT and MWCNT), each with distinct thermophysical properties outlined in Table 2. The assumptions governing the flow include incompressibility, steadiness, laminarity, and single-phase behavior of the nanofluid. Parameters associated with radiation and viscous dissipation are neglected in this study.

### 2.2. Governing equations and boundary conditions

The governing equations for the above two-dimensional problem can be formulated by adhering to principles of mass, momentum, and energy conservation. The flow is characterized by being laminar, incompressible, and in a steady state [1,65].

$$\nabla \cdot \mathbf{u} = 0 \tag{1}$$

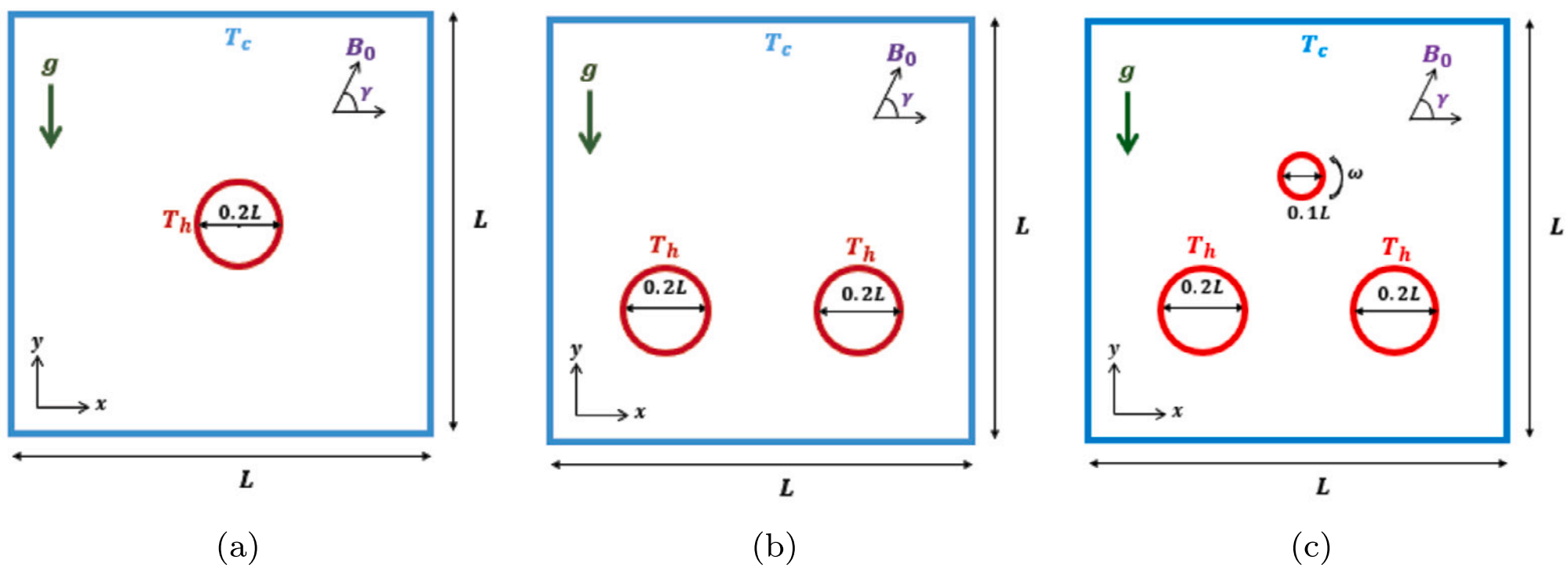

**Fig. 1.** Schematic diagram of the physical problem: (a) **Case 1** – one heated cylinder in a square enclosure, (b) **Case 2** – two heated cylinders in a square enclosure, and (c) **Case 3** – two heated cylinders in a square enclosure with a flow deflector.

**Table 1**
Description of non-dimensional boundary conditions.

| Boundaries | Velocity condition | Thermal condition |
|---|---|---|
| Cavity wall | $U = V = 0$ | $\theta = 0$ |
| Heated cylinder wall | $U = V = 0$ | $\theta = 1$ |
| Flow deflector wall | $\omega$ | $\partial\theta/\partial X = \partial\theta/\partial Y = 0$ |
| Supportive insulator wall | $U = V = 0$ | $\partial\theta/\partial X = \partial\theta/\partial Y = 0$ |

**Table 2**
Thermo-physical properties of pure water and CNT nanoparticles at T = 25 °C (SI units). [1,18].

| Particles | $C_p$ | $\rho$ | $K$ | $\beta$ | $\mu$ | $\sigma$ |
|---|---|---|---|---|---|---|
| Pure Water | 4179 | 997.1 | 0.613 | $21\times10^{-5}$ | $8.9\times10^{4}$ | 0.05 |
| SWCNT | 425 | 2600 | 6600 | $1.9\times10^{-5}$ | – | $2.54\times10^{4}$ |
| MWCNT | 796 | 1600 | 3000 | $2.1\times10^{-5}$ | – | $4.95\times10^{3}$ |

$$\mathbf{u}\cdot\nabla\mathbf{u} = -\frac{1}{\rho_{nf}}\nabla P + \nu_{nf}\nabla^2\mathbf{u} + g\beta_{nf}(T - T_c) + \frac{1}{\rho_{nf}}\mathbf{J}\times\mathbf{B} \tag{2}$$

$$\mathbf{u}\cdot\nabla T = \alpha_{nf}\nabla^2 T \tag{3}$$

The dimensionless form of the above governing equations can be obtained by considering the following parameters:

$$X, Y = \frac{x, y}{L};\quad U, V = \frac{u, v}{V_0};\quad \theta = \frac{T - T_c}{T_h - T_c};\quad P = \frac{p}{\rho_{nf}V_0^2};\quad Re = \frac{V_0 L}{\nu_f};$$

$$Ha = B_0 L\sqrt{\frac{\sigma_f}{\mu_f}};\quad \text{and}\quad Ri = \frac{Ra}{Pr\cdot Re^2} \tag{4}$$

The dimensionless parameters are chosen to characterize key transport mechanisms in mixed convection. The Reynolds number ($Re$) governs convective transport, the Richardson number ($Ri$) balances natural and forced convection, and the Hartmann number ($Ha$) quantifies magnetic field effects on flow stability. Along with nanoparticle volume fraction ($\phi$) and geometric variations, these parameters ensure a comprehensive evaluation of heat transfer and entropy generation. Using these, the dimensionless forms for the conservation of mass, momentum, and energy ((1)–(3)) for the nanofluid flow can be written as [1,55]:

$$\frac{\partial U}{\partial X} + \frac{\partial V}{\partial Y} = 0 \tag{5}$$

$$\frac{\partial U}{\partial X} + \frac{\partial V}{\partial Y} = 0$$

$$U\frac{\partial U}{\partial X} + V\frac{\partial U}{\partial Y} = -\frac{\partial P}{\partial X} + \frac{\nu_{nf}}{\nu_f}\cdot\frac{1}{Re}\left(\frac{\partial^2 U}{\partial X^2} + \frac{\partial^2 U}{\partial Y^2}\right) + \frac{Ha^2}{Re}\frac{\rho_f}{\rho_{nf}}\frac{\sigma_{nf}}{\sigma_f}\left(V\sin\gamma\cos\gamma - U\sin^2\gamma\right) \tag{6}$$

$$U\frac{\partial V}{\partial X} + V\frac{\partial V}{\partial Y} = -\frac{\partial P}{\partial Y} + \frac{\nu_{nf}}{\nu_f}\cdot\frac{1}{Re}\left(\frac{\partial^2 V}{\partial X^2} + \frac{\partial^2 V}{\partial Y^2}\right) + \frac{Ha^2}{Re}\frac{\rho_f}{\rho_{nf}}\frac{\sigma_{nf}}{\sigma_f}\left(U\sin\gamma\cos\gamma - V\cos^2\gamma\right) + \frac{\beta_{nf}}{\beta_f}Ri\,\theta \tag{7}$$

$$U\frac{\partial\theta}{\partial X} + V\frac{\partial\theta}{\partial Y} = \frac{\alpha_{nf}}{\alpha_f}\cdot\frac{1}{Re\,Pr}\left(\frac{\partial^2\theta}{\partial X^2} + \frac{\partial^2\theta}{\partial Y^2}\right) \tag{8}$$

Along with the non-dimensional equations (5)–(8), the boundary conditions listed in Table 1 define the velocity and thermal constraints for different surfaces. The flow deflector rotates with a velocity $\omega$, enforcing a no-slip condition while maintaining thermal insulation ($\partial\theta/\partial X = \partial\theta/\partial Y = 0$), ensuring no additional heat flux contribution to the system.

### 2.3. Thermophysical properties of nanofluids

The effective thermophysical properties of the nanofluid in the above equations are calculated using the following equations:

$$\rho_{nf} = \phi\rho_s + (1-\phi)\rho_f \tag{9}$$

$$(\rho C_p)_{nf} = \phi(\rho C_p)_s + (1-\phi)(\rho C_p)_f \tag{10}$$

$$\alpha_{nf} = \frac{k_{nf}}{(\rho C_p)_{nf}} \tag{11}$$

$$(\rho\beta)_{nf} = \phi(\rho\beta)_s + (1-\phi)(\rho\beta)_f \tag{12}$$

The following equations are used to calculate the effective thermal and electrical conductivity of nanoparticles in liquid, as presented by Maxwell [56]:

$$\frac{k_{nf}}{k_f} = \frac{k_s + 2k_f - 2\phi(k_f - k_s)}{k_s + 2k_f + \phi(k_f - k_s)} \tag{13}$$

$$\frac{\sigma_{nf}}{\sigma_f} = 1 + \frac{3(\xi - 1)\phi}{(\xi + 2) - (\xi - 1)\phi} \tag{14}$$

where $\xi = \sigma_s/\sigma_f$.

Brinkman presented the following well-validated model for calculating the viscosity of nanofluid considering the volume fraction of the nanoparticles [57]:

$$\mu_{nf} = \mu_f(1-\phi)^{-2.5} \tag{15}$$

### 2.4. Output parameters

The local Nusselt number is obtained by performing an energy balance on the heated cylinder surface as:

$$Nu_l = -\frac{k_{nf}}{k_f}\left(\frac{\partial\theta}{\partial X} + \frac{\partial\theta}{\partial Y}\right) \tag{16}$$

**Table 3**
Grid sensitivity test based on average Nusselt number calculation at the heated surfaces using SWCNT nanofluids. Significant values are indicated in bold.

| Grid | Element | | | $Nu_{ave}$ | | |
|---|---|---|---|---|---|---|
| | Type | Number | Maximum Size | Case 1 | Case 2 | Case 3 |
| G1 | Triangular | 20504 | 0.0200 | 5.5935 | 9.8110 | 9.4049 |
| **G2** | | **35786** | **0.0100** | **5.5924** | **9.8097** | **9.4039** |
| G3 | | 50310 | 0.0080 | 5.5915 | 9.8085 | 9.4027 |
| G4 | | 73810 | 0.0060 | 5.5896 | 9.8057 | 9.4003 |
| G5 | Quadrilateral | 15349 | 0.0100 | 5.6064 | 9.8414 | 9.3997 |
| G6 | | 22544 | 0.0080 | 5.6066 | 9.8399 | 9.4031 |
| G7 | | 38422 | 0.0060 | 5.6120 | 9.8487 | 9.4112 |
| G8 | | 78306 | 0.0040 | 5.6144 | 9.8523 | 9.4235 |

The entropy generation due to the non-equilibrium flow characteristics in the domain can be calculated using the following formula [55]:

$$S_t = \underbrace{\left(\frac{k_{nf}}{k_f}\right)\left[\left(\frac{\partial\theta}{\partial X}\right)^2 + \left(\frac{\partial\theta}{\partial Y}\right)^2\right]}_{S_h} \underbrace{+\,\theta\cdot\left(\frac{\mu_{nf}}{\mu_f}\right) Re^2\cdot Pr_2\left\{2\left[\left(\frac{\partial U}{\partial X}\right)^2 + \left(\frac{\partial V}{\partial Y}\right)^2\right] + \left(\frac{\partial V}{\partial X} + \frac{\partial U}{\partial Y}\right)^2\right\}}_{S_v} \underbrace{+\,\theta\cdot\left(\frac{\sigma_{nf}}{\sigma_f}\right) Re^2\cdot Pr_2\cdot Ha^2\cdot(U\sin\gamma - V\cos\gamma)^2}_{S_j} \quad (17)$$

In this context, the non-dimensional local entropy generation rates attributed to heat conduction, fluid friction, and Joule heating are represented by $S_h$, $S_v$, and $S_j$, respectively. The ratio between viscous and thermal entropy generation, termed as the irreversibility factor $\Theta$, is set to a constant value of $10^{-4}$.

The local Bejan number is defined as the fraction of entropy generation due to heat transfer relative to the total entropy generation and can be expressed as follows [55,58]:

$$Be_l = \frac{S_h}{S_t} \quad (18)$$

The average properties, $P_{ave}$, can be obtained by integrating the local value, $P_l$, over the surface, $A$, of the heated cylinders as follows:

$$P_{ave} = \frac{1}{A}\int_0^A P_l\, dA \quad (19)$$

## 3. Numerical methods

The thermal performance study of the innovative enclosure design, outlined in Section 2, is conducted using the COMSOL Multiphysics 5.5 software (https://www.comsol.com/release/5.5). The solution methodology involves employing the Galerkin weighted residual Finite Element Method (GFEM) to solve the governing equations (5)–(8) at designated mesh points, facilitated by the incorporation of specified boundary conditions defined in Table 1.

To ensure the adequacy of the mesh arrangement and validate the independence of the findings from the grid, grid sensitivity tests were performed, focusing on the calculation of the average Nusselt number as detailed in Table 3. These simulations confirmed that refining the grid beyond a certain resolution did not significantly alter the results, ensuring that the selected grid adequately captures the physics of the problem. The calculation of the average Nusselt number involved setting $Re = 100$, $Pr = 6.96$, $Ri = 1$, $Ha = 20$, $\gamma = 30°$, $\phi = 0.06$, and the deflector used in **Case 3** with rotational speed, $\omega$ set at 1. The chosen grid distribution is illustrated in Fig. 2.

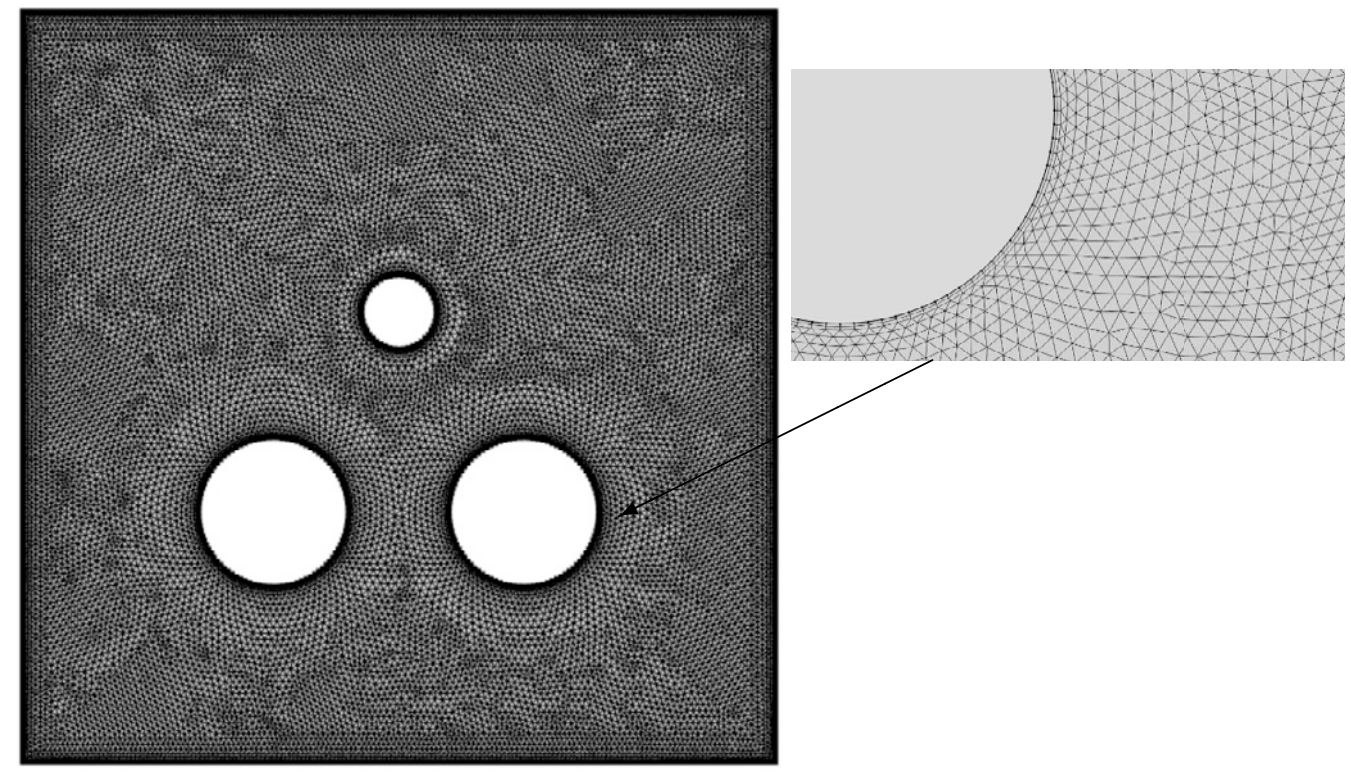

**Fig. 2.** Mesh structures with a zoomed-in region for clarity (Case 3).

We examined a lid-driven square cavity with a heated bottom wall and a cooled right wall, while the top wall remained adiabatic and the left wall was linearly heated, to confirm our computational setup [59]. An excellent agreement was obtained in terms of streamlines and isotherms, as well as the average Nusselt number calculation, as shown in Fig. 3.

## 4. Results and discussions

As discussed above, this manuscript investigates the effect of mixed convection of CNT-water nanofluid in a cavity with cylinders under the influence of a magnetic field. Three cases are considered, as illustrated in Fig. 1, focusing on the number of cylinders and the position of the flow deflector. In each case, the cylinders are heated while the boundary walls are cold, and the circular flow deflector is insulated and rotates with a unit angular speed. Specifically, **Case 1** places a single cylinder in the middle of the cavity. **Case 2** introduces a second heated cylinder, creating more complex thermal and flow interactions. **Case 3** further modifies the setup by adding a flow deflector, resulting in additional flow complexity and enhanced fluid mixing. The results for these three configurations are presented and analyzed.

Table 4 outlines the parameters based on fluid characteristics, external forces, and geometric shapes. The subsequent sections detail the observed flow and thermal patterns, entropy generation, and Bejan number variations, providing insights into the influence of geometry and magnetic fields on the thermal and fluid behavior of the nanofluid system.

**Case 1:** The effects of varying the Richardson number ($Ri$), Hartmann number ($Ha$), and wavy wall peaks ($n$) on streamlines, isotherms, local entropy generation ($S_l$), and the local Bejan number ($Be_l$) for **Case 1** are illustrated in Fig. 4, Fig. 5, and Fig. 6, respectively. As $Ri$ increases from 0.01 to 10, streamlines the transition from a symmetric double-vortex structure to a more complex and asymmetric configuration, indicating the increasing dominance of buoyancy forces. The isotherms show increasing stratification and the formation of a thermal boundary layer near the top of the cavity at higher $Ri$, reflecting enhanced thermal buoyancy effects. Local entropy generation ($S_l$) initially appears symmetrically near the cylinder at low $Ri$ but shifts to higher values near the heated walls with increasing $Ri$, indicating greater irreversibilities due to heat transfer and fluid friction. The local Bejan number ($Be_l$) distributions reveal that thermal entropy generation becomes more dominant, especially near the hot walls as $Ri$ increases, highlighting the significant influence of thermal effects on overall entropy generation. Similarly, increasing the Reynolds number ($Re$) produces comparable results, as it enhances inertial forces relative to viscous forces, aligning with the relationship, $Gr = Re^2 \cdot Ri$.

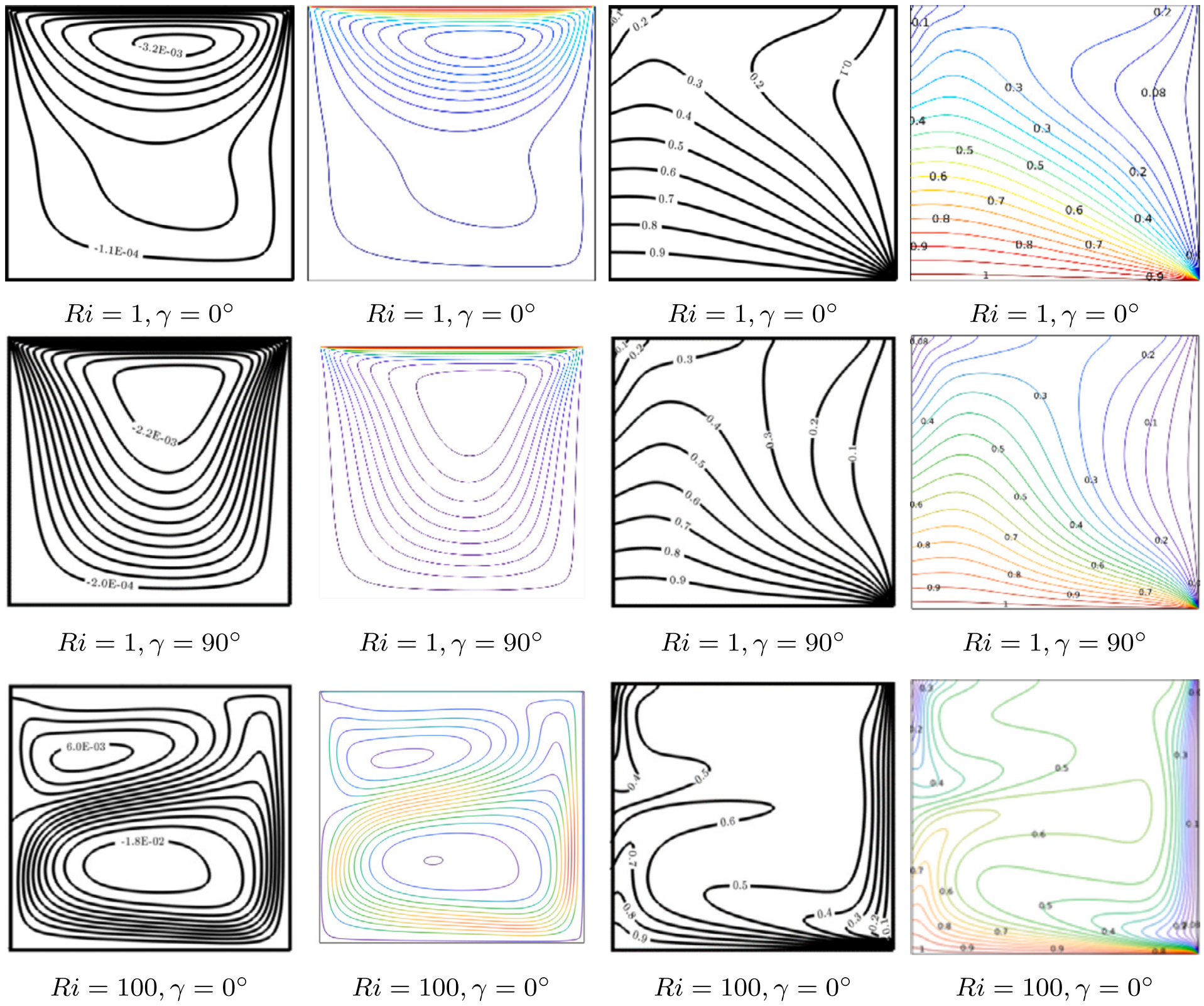

**Fig. 3.** Model validation study with streamline variation and isothermal contours inside a square cavity with $Re = 100$, $Pr = 0.71$, and $Ha = 25$.

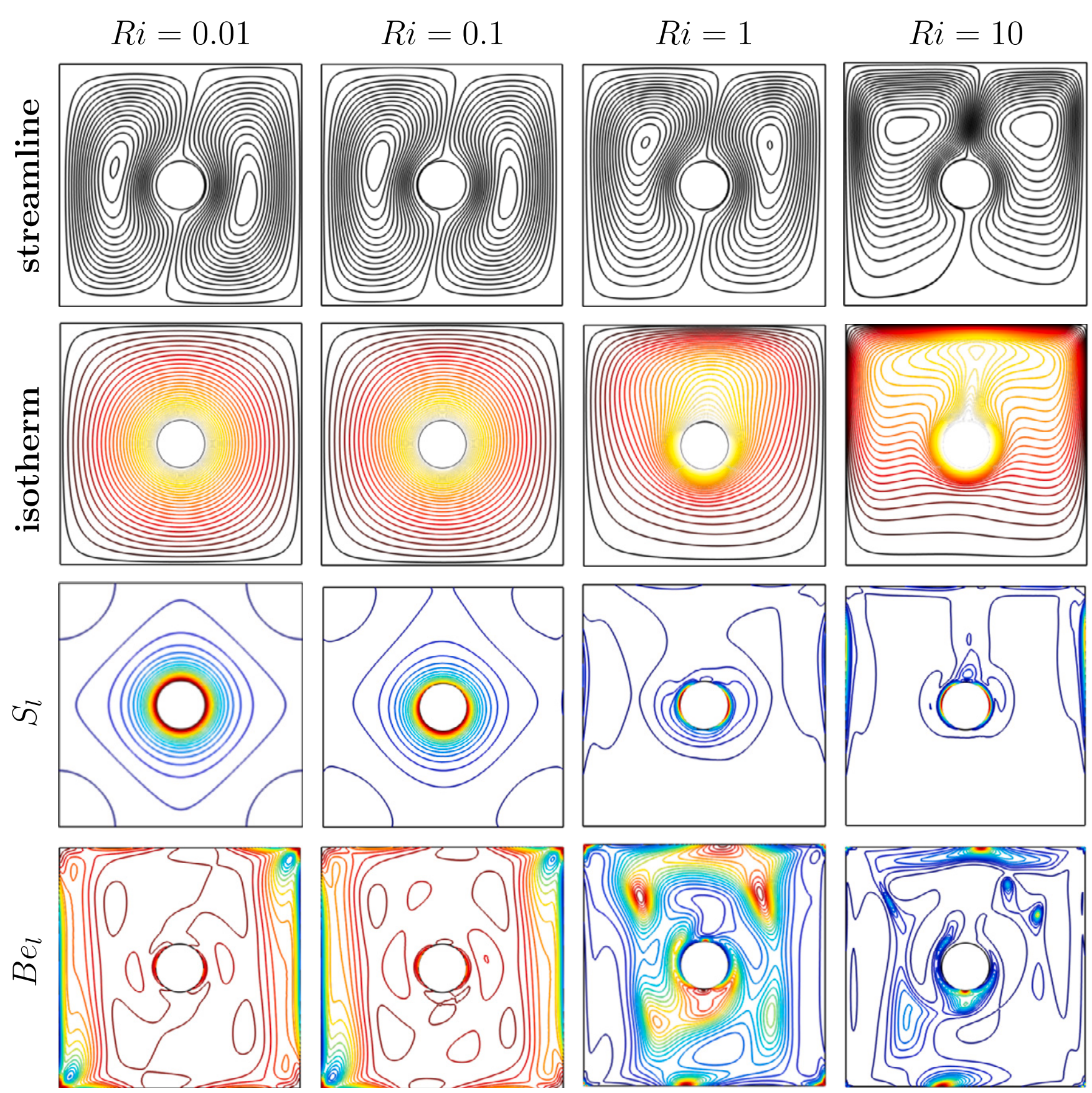

**Fig. 4.** Effects of Richardson number ($Ri$) variations on streamlines, isotherms, local entropy generation, and local Bejan number when $Re = 100$, $Ha = 20$, $\phi = 0.06$, and $\gamma = 30°$ for **Case 1**.

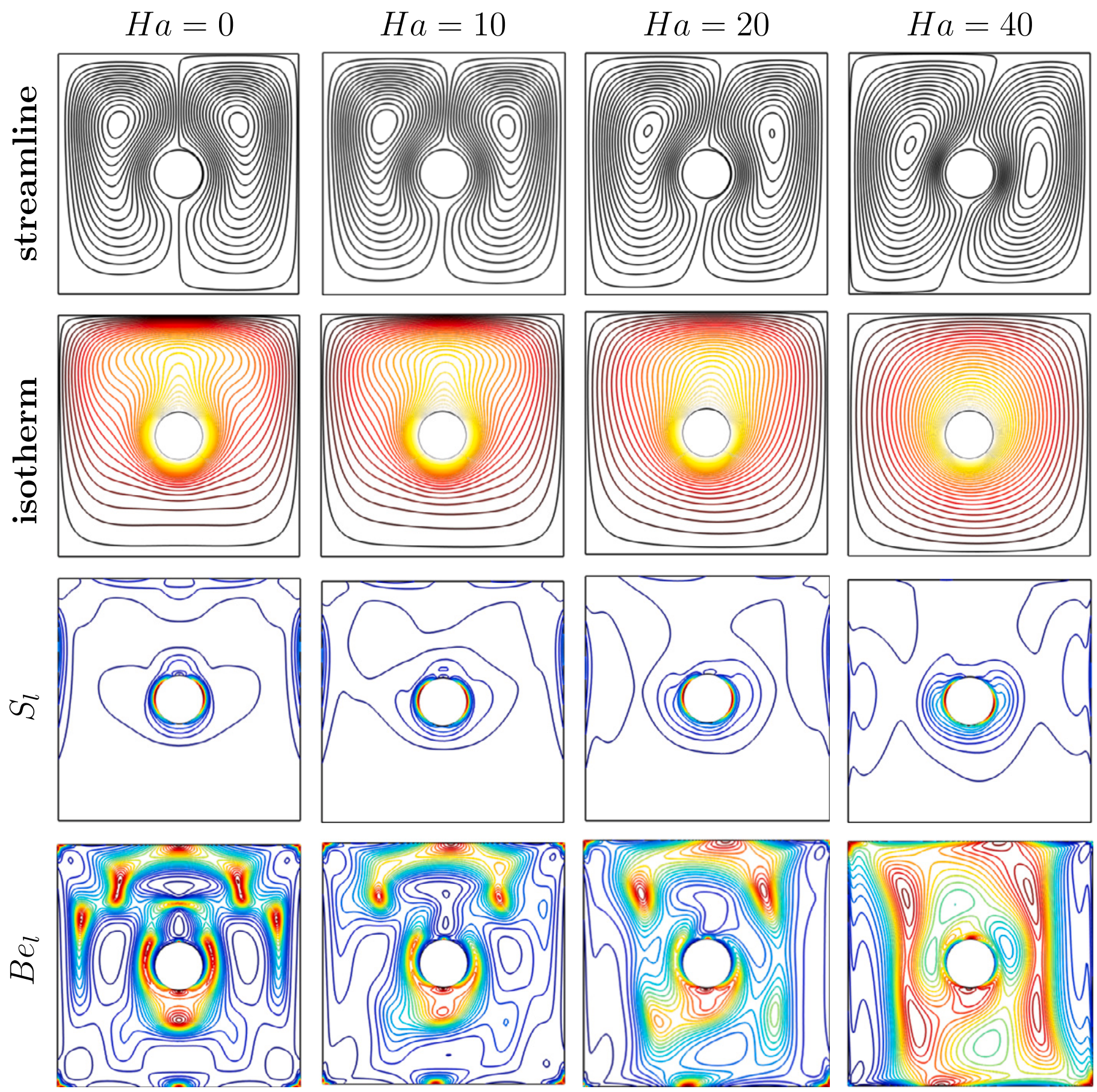

**Fig. 5.** Effects of Hartmann number ($Ha$) variations on streamlines, isotherms, local entropy generation, and local Bejan number when $Re = 100$, $Ri = 1$, $\phi = 0.06$, and $\gamma = 30°$ for **Case 1**.

**Table 4**
Parameters studied based on fluid characteristics, external forces, and geometric shapes.

| Parameters | Ranges | SWCNT | MWCNT | Case 1 | Case 2 | Case 3 |
|---|---|---|---|---|---|---|
| Reynolds number ($Re$) | 1, 10, 50, 100, 200 | ✓ | ✓ | ✓ | ✓ | ✓ |
| Richardson number ($Ri$) | 0.01, 0.1, 1, 10 | ✓ | ✓ | ✓ | ✓ | ✓ |
| Hartman number ($Ha$) | 0, 10, 20, 40 | ✓ | ✓ | ✓ | ✓ | ✓ |
| Inclination angle ($\alpha$) | 0, 30, 45, 60, 90 | ✓ | ✓ | ✓ | ✓ | ✓ |
| Rotational Speed ($\omega$) | 1, 3, 5, 7 | ✓ | ✓ | ✓ | ✓ | ✓ |
| Volume fraction ($\phi$) | 0, 0.02, 0.04, 0.06, 0.08 | ✓ | ✓ | ✓ | ✓ | ✓ |
| Wavy wall peaks ($n$) | 0, 1, 2, 3, 4 | ✓ | ✓ | ✓ | ✓ | ✓ |
| Hartman angle ($\gamma$) | 0, 30, 45, 60, 90 | ✓ | ✓ | ✓ | ✓ | ✓ |

As $Ha$ increases from 0 to 40, the streamlines indicate a suppression of vortices and more streamlined flow due to the increasing influence of the magnetic field. The isotherms become more uniformly distributed with higher $Ha$, reflecting the suppression of convective currents by the magnetic field. The local entropy generation ($S_l$) decreases around the central cylinder with increasing $Ha$, suggesting reduced irreversibilities. The local Bejan number ($Be_l$) shows that thermal effects continue to dominate over viscous effects, with higher $Ha$ leading to a more uniform distribution of $Be_l$. These findings highlight the significant role of magnetic fields in stabilizing flow patterns and enhancing thermal uniformity. When examining the impact of wavy wall peaks ($n$), the streamlines exhibit more complex patterns as $n$ increases, indicating enhanced fluid mixing due to the wavy walls. The isotherms show that higher $n$ leads to more irregular thermal distributions, reflecting increased convective heat transfer. Entropy generation is concentrated around the cylinder and near the wavy walls, with higher $n$ resulting in more dispersed regions of $S_l$. The Bejan number contours indicate a nuanced balance between heat transfer and fluid friction, with higher $n$ enhancing fluid friction effects due to the increased surface area of the wavy walls. These findings underscore the importance of wall geometry in enhancing fluid mixing and heat transfer efficiency in thermal systems.

The local entropy generation contours for **Case 1** are illustrated in Fig. 7, where entropy generation due to heat transfer and fluid friction is concentrated around the heated cylinder, with Joule heating effects remaining highly localized. The Bejan number distribution indicates that heat transfer dominates near the walls, while fluid friction is significant around the object.

The average Nusselt number variations for **Case 1** due to the volume fraction of nanoparticles ($\phi$), Hartmann angle ($\gamma$), and wavy wall peaks ($n$) are presented in Fig. 8. Increasing the volume fraction of nanoparticles enhances the thermal conductivity of the nanofluid, leading to higher Nusselt numbers, as observed in Fig. 8a. This effect is more pronounced at higher Reynolds numbers, indicating the synergistic effect of nanoparticle dispersion and fluid inertia. Fig. 8b reveals that increasing the Hartmann angle generally improves heat transfer by stabilizing the flow, with the highest Nusselt numbers occurring at a Hartmann angle of 30°. Fig. 8c shows that wavy wall peaks induce more complex flow structures, enhancing mixing and, consequently, heat transfer. The combination of higher Reynolds numbers and wavy wall geometries results in the most significant improvements in the Nusselt number, demonstrating the critical role of wall geometry in thermal management.

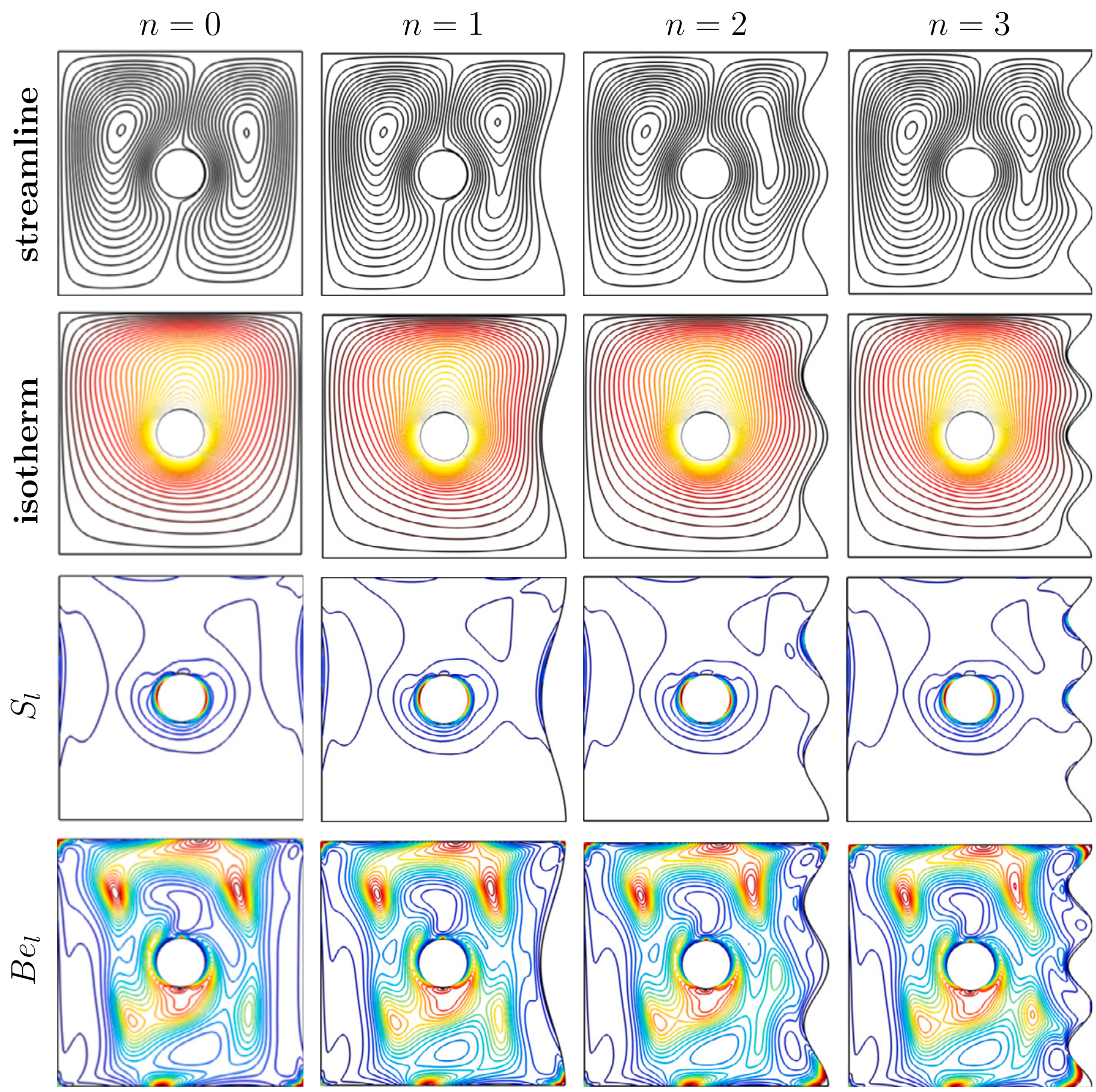

**Fig. 6.** Effects of wavy wall peaks ($n$) variations on streamlines, isotherms, local entropy generation, and local Bejan number when $Re = 100$, $Ri = 1$, $Ha = 20$, $\phi = 0.06$, and $\gamma = 30^\circ$ for **Case 1**.

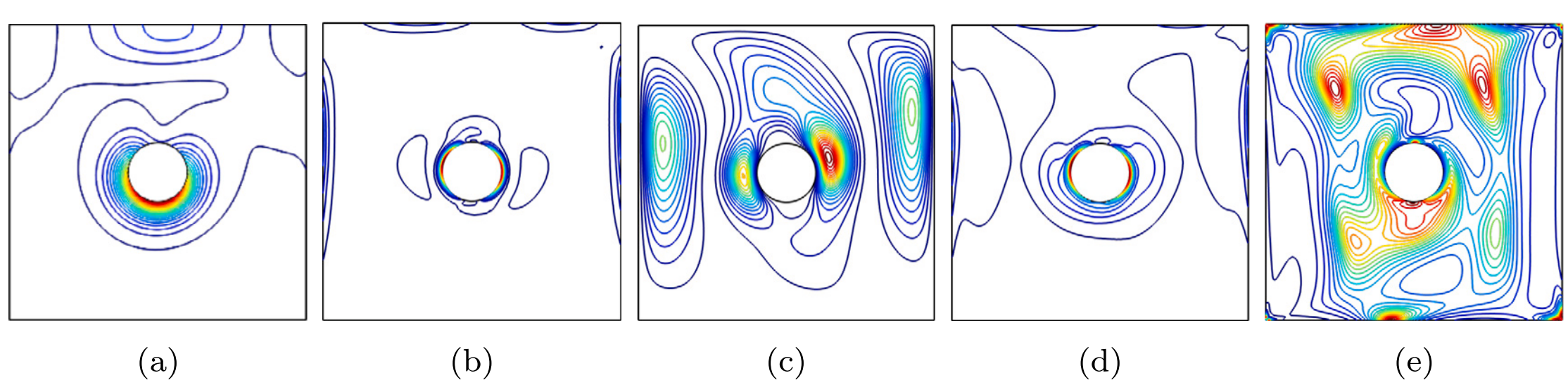

**Fig. 7.** Local entropy generation contours when $Re = 100$, $Ri = 1$, $Ha = 20$, $\phi = 0.06$, and $\gamma = 30^\circ$ for **Case 1**; (a) local entropy generation due to heat transfer, (b) local entropy generation due to fluid friction, (c) local entropy generation due to Joule heating, (d) local entropy generation, (e) local Bejan number.

Fig. 9 shows that the average entropy generation ($S_t$) increases with the Reynolds number ($Re$), particularly for higher irreversibility factors ($\Theta$), reflecting increased energy dissipation due to more vigorous fluid motion. The average Bejan number ($Be_{\text{ave}}$) decreases gradually with increasing $Re$, indicating a shifting contribution of heat transfer and fluid friction to entropy generation.

**Case 2:** The impacts of varying the Richardson number ($Ri$) on streamlines, isotherms, local entropy generation ($S_l$), and the local Bejan number ($Be_l$) for **Case 2**, featuring two heated cylinders, are shown in Fig. 10. The streamlines reveal dual vortices around each cylinder, becoming more pronounced with increasing $Ri$. Isotherms indicate strong thermal interactions between the cylinders, with merging thermal boundary layers at higher $Ri$, signifying enhanced heat transfer. Entropy generation spreads between the cylinders and near the walls, while the Bejan number distribution suggests an increased influence of fluid friction on entropy generation as $Ri$ rises. Compared to **Case 1**, the presence of a second cylinder leads to more complex flow and thermal interactions, resulting in higher overall entropy generation and more intricate patterns of $Be_l$.

As shown in Fig. 11, increasing the Hartmann number ($Ha$) further modifies these parameters. The streamlines indicate that higher $Ha$ suppresses the vortices around each cylinder, leading to more stable and streamlined flow patterns. This magnetic damping effect results in more uniform thermal gradients between the cylinders, as seen in the isotherms, indicating reduced convective heat transfer. With higher $Ha$, entropy generation becomes more localized, particularly between the cylinders and near the walls. The Bejan number contours suggest an increasing dominance of fluid friction in entropy generation, similar to **Case 1**, but with more complex interactions due to the dual-cylinder configuration. These observations highlight the magnetic field's role in stabilizing fluid motion and enhancing thermal management.

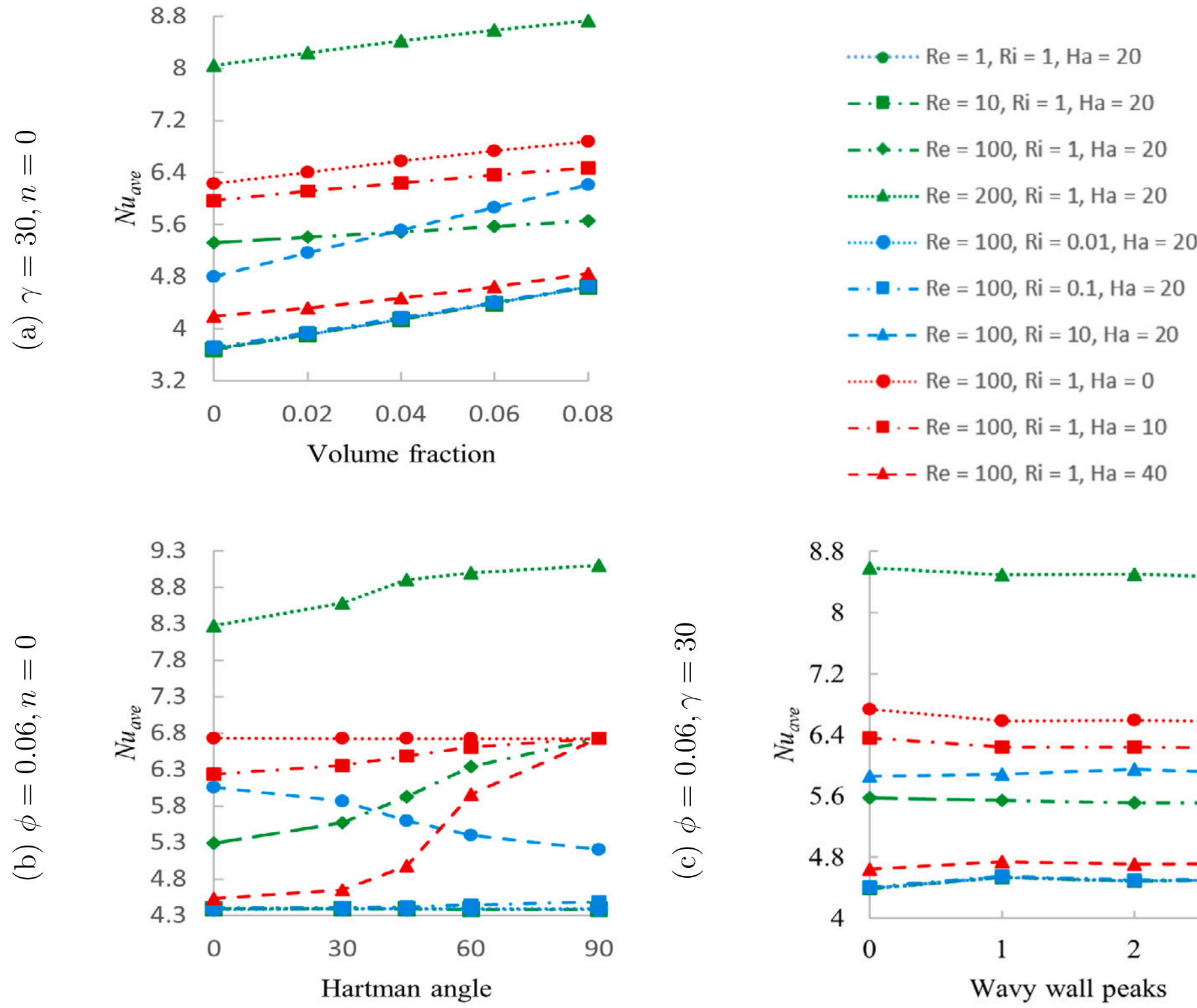

**Fig. 8.** Average Nusselt number variations for **Case 1** due to effects of (a) volume fraction of nanoparticles, $\phi$; (b) Hartmann angle, $\gamma$; and (c) wavy wall peaks, $n$.

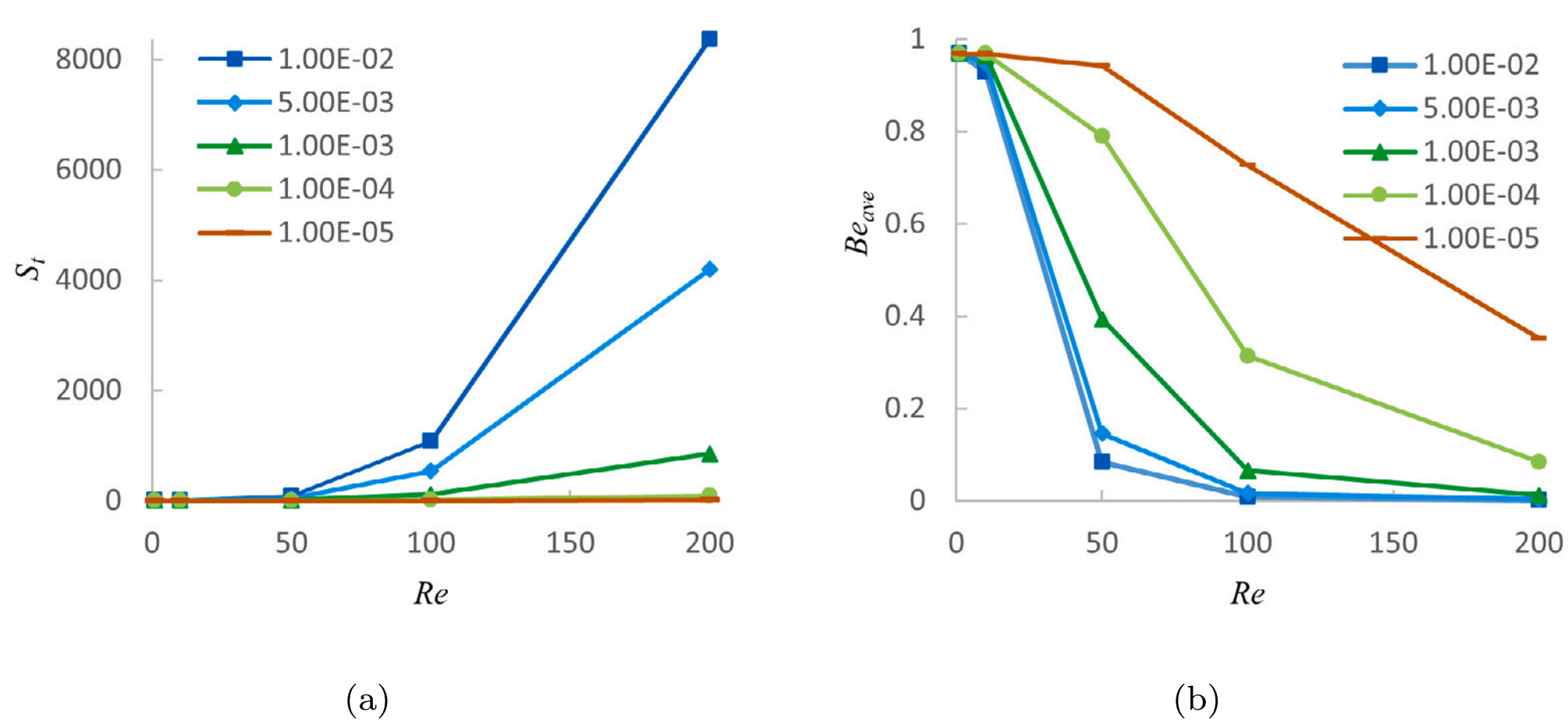

**Fig. 9.** Effects of irreversibility factor, Θ for **Case 1** on (a) average entropy generation, $S_t$; (b) average Bejan number, $Be_{ave}$.

The effects of wavy wall peaks ($n$) on the same parameters are depicted in Fig. 12. The introduction of wavy walls causes the streamlines to exhibit more intricate vortex structures around each cylinder, which become increasingly complex as $n$ rises, indicating stronger fluid interactions and enhanced mixing. The isotherms show that higher $n$ results in more irregular thermal gradients, enhancing heat transfer efficiency between the cylinders. Entropy generation spreads more broadly with higher $n$, concentrating around the cylinders and wavy walls. The Bejan number distribution reveals a heightened influence of fluid friction on entropy generation with increasing $n$, demonstrating greater thermal and fluid dynamic interactions due to the wavy wall geometry. These findings emphasize the impact of wall geometry on improving fluid mixing and thermal performance in multi-cylinder systems.

Fig. 13 depicts the local entropy generation contours for **Case 2**, showing similar patterns to **Case 1** but with increased intensity and spread, indicating higher thermal gradients and velocity effects. The Bejan number indicates a greater influence of fluid friction in more regions than in **Case 1**.

Fig. 14 presents the average Nusselt number variations for **Case 2**, focusing on the impacts of nanoparticle volume fraction ($\phi$), Hartmann angle ($\gamma$), and wavy wall peaks ($n$). As seen in Fig. 14a, the inclusion of nanoparticles enhances thermal performance, with the effect becoming more prominent at higher Reynolds numbers. In Fig. 14b, increasing the Hartmann angle initially improves the Nusselt number due to the stabilizing effect of the magnetic field on the flow. However, at very high $\gamma$ values, the Nusselt number stabilizes, indicating a balance between magnetic damping and convective enhancement. Fig. 14c highlights that wavy wall peaks significantly improve heat transfer, especially at higher Reynolds numbers and Hartmann angles, by promoting turbulence and enhancing thermal mixing. Compared to **Case 1**, the presence of two cylinders in **Case 2** introduces additional thermal interactions, resulting in overall higher Nusselt numbers and more pronounced effects of the Hartmann angle and wavy walls.

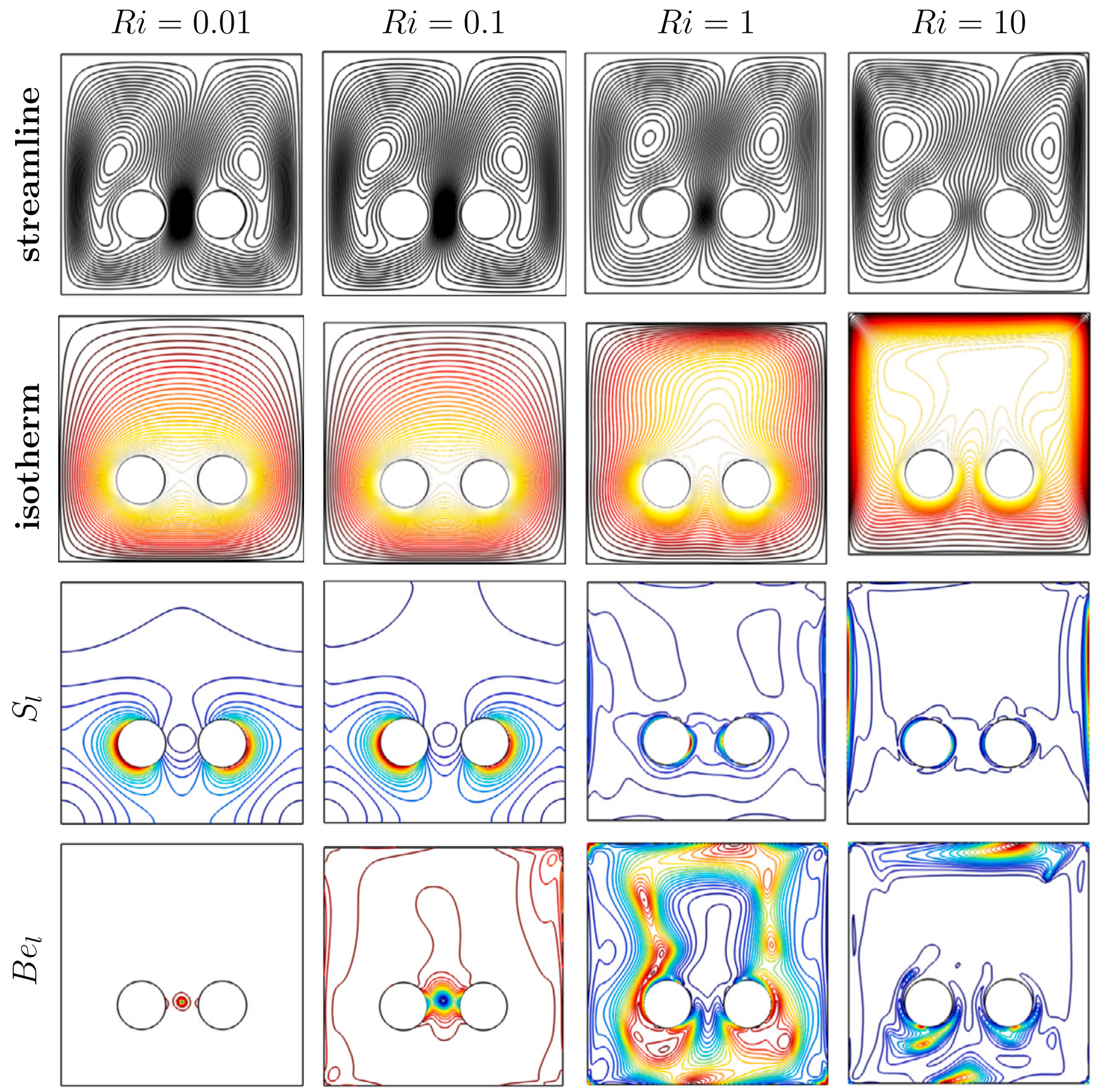

**Fig. 10.** Effects of Richardson number ($Ri$) variations on streamlines, isotherms, local entropy generation, and local Bejan number when $Re = 100$, $Ha = 20$, $\phi = 0.06$, and $\gamma = 30°$ for **Case 2**.

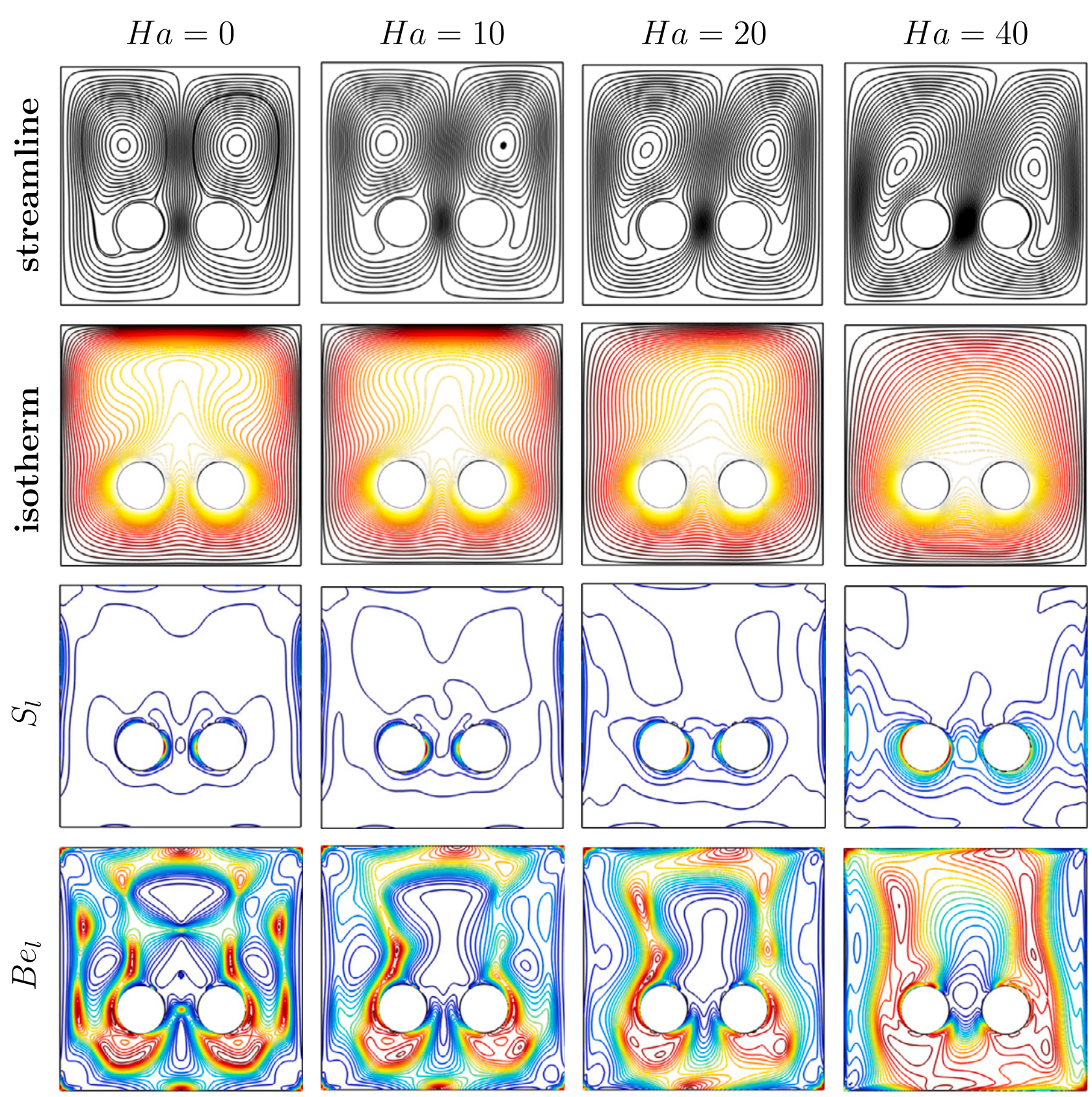

**Fig. 11.** Effects of Hartmann number ($Ha$) variations on streamlines, isotherms, local entropy generation, and local Bejan number when $Re = 100$, $Ri = 1$, $\phi = 0.06$, and $\gamma = 30°$ for **Case 2**.

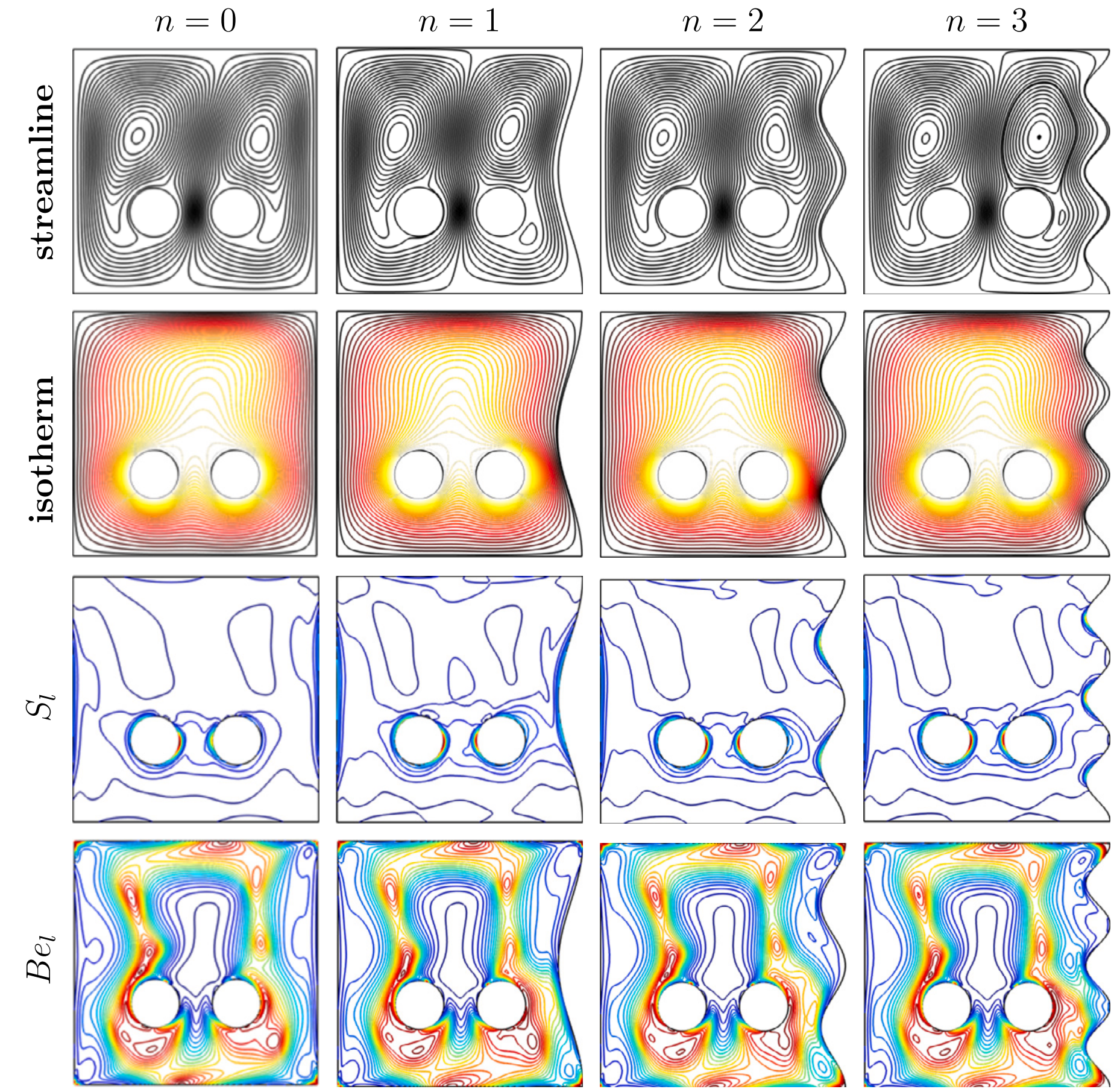

**Fig. 12.** Effects of wavy wall peaks ($n$) variations on streamlines, isotherms, local entropy generation, and local Bejan number when $Re = 100$, $Ri = 1$, $Ha = 20$, $\phi = 0.06$, and $\gamma = 30°$ for **Case 2**.

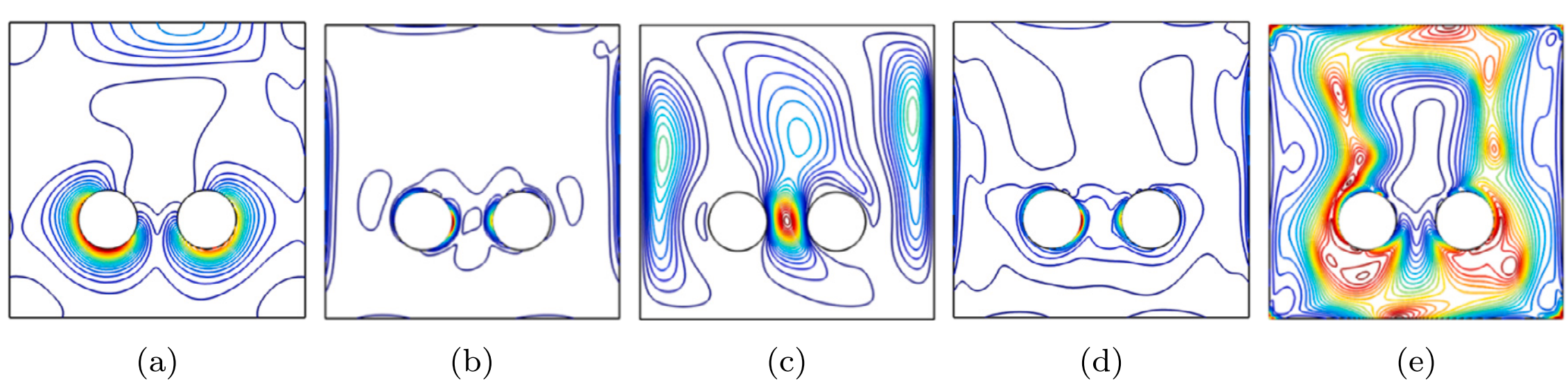

**Fig. 13.** Local entropy generation contours when $Re = 100$, $Ri = 1$, $Ha = 20$, $\phi = 0.06$, and $\gamma = 30°$ for **Case 2**; (a) local entropy generation due to heat transfer, (b) local entropy generation due to fluid friction, (c) local entropy generation due to Joule heating, (d) local entropy generation, (e) local Bejan number.

The entropy generation ($S_t$) patterns are intensified and more widespread for **Case 2** compared to **Case 1**, as illustrated in Fig. 15, with a significant increase at higher $Re$ and $\Theta$. The average Bejan number ($Be_{\text{ave}}$) shows a steep decline with increasing $Re$, suggesting that fluid friction plays a greater role in the entropy generation process in the presence of two heated cylinders.

**Case 3:** Fig. 16 examines the effects of varying the Richardson number ($Ri$) on streamlines, isotherms, local entropy generation ($S_l$), and the local Bejan number ($Be_l$) for **Case 3**, which incorporates a flow deflector with two heated cylinders. The flow deflector significantly alters streamline patterns, creating complex vortex structures, especially at higher $Ri$, indicating intensified mixing and flow disruption. Isotherms are asymmetrically distributed due to the deflector, with stronger thermal gradients observed, leading to more efficient heat transfer in localized regions. Entropy generation is highest around the cylinders and the deflector, with $S_l$ showing more dispersed and extensive regions of energy dissipation. The $Be_l$ contours indicate a nuanced balance between heat transfer and fluid friction, showcasing the deflector's role in enhancing entropy generation dominated by fluid friction at elevated $Ri$. Comparing **Case 3** with **Case 2**, the deflector introduces additional complexity and enhances fluid friction effects, demonstrating the significant impact of geometric modifications on convection and entropy generation, suggesting potentially more effective thermal management strategies.

The impact of increasing the Hartmann number ($Ha$) on the same parameters is shown in Fig. 17. The flow deflector adds complexity, with streamlines indicating suppressed vortex structures at higher $Ha$, signifying strong magnetic damping effects. Isotherms display more symmetric thermal distributions due to the deflector's influence, particularly at higher $Ha$. Entropy generation peaks around the deflector and the cylinders, with $S_l$ becoming more localized as $Ha$ increases. The Bejan number contours highlight a nuanced balance between heat transfer and fluid friction, with $Ha$ enhancing the dominance of entropy generation driven by fluid friction, especially around the deflector. These findings underscore the importance of geometric modifications and magnetic fields in optimizing thermal systems by altering flow and thermal fields.

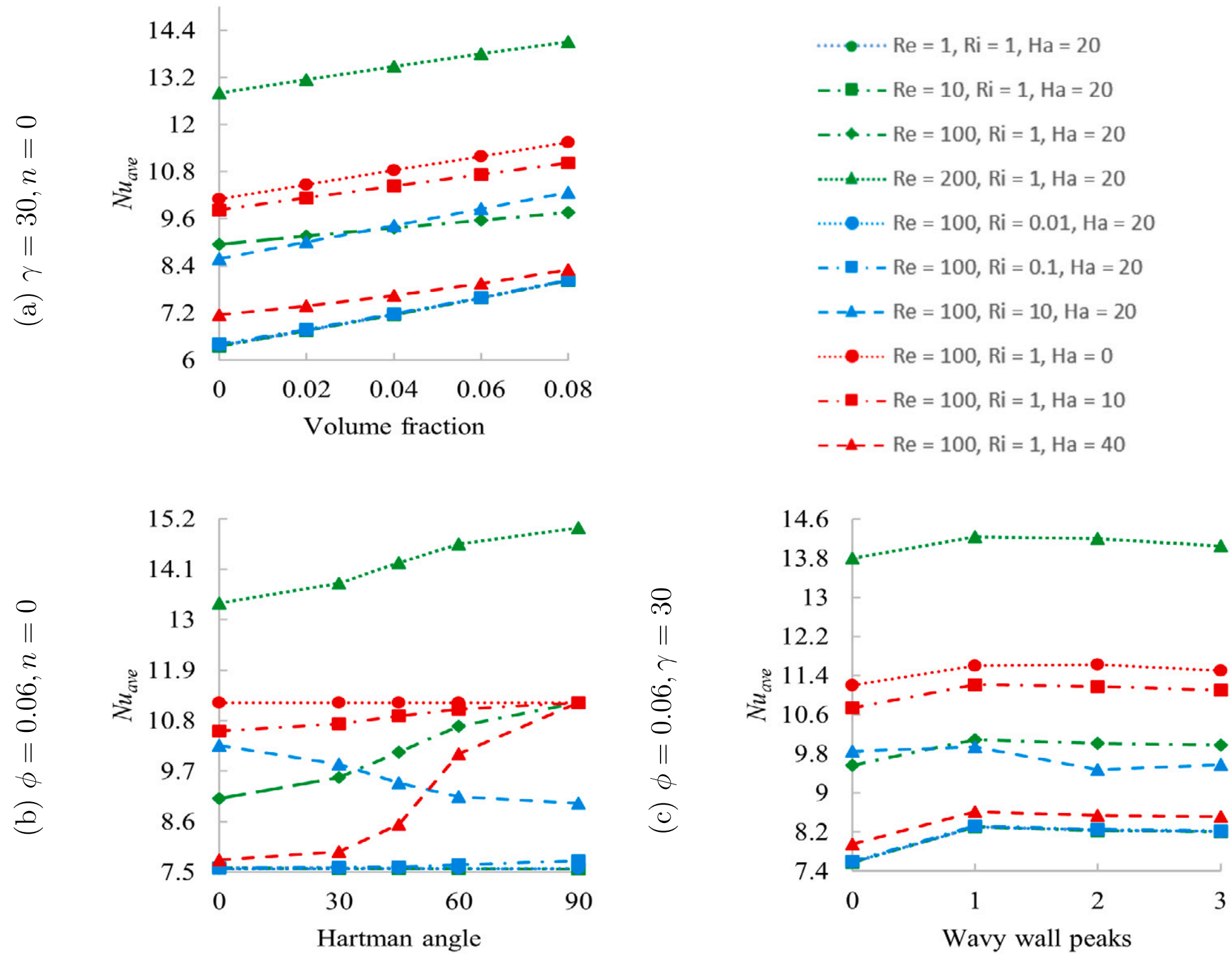

**Fig. 14.** Average Nusselt number variations for **Case 2** due to effects of (a) volume fraction of nanoparticles, $\phi$; (b) Hartmann angle, $\gamma$; and (c) wavy wall peaks, $n$.

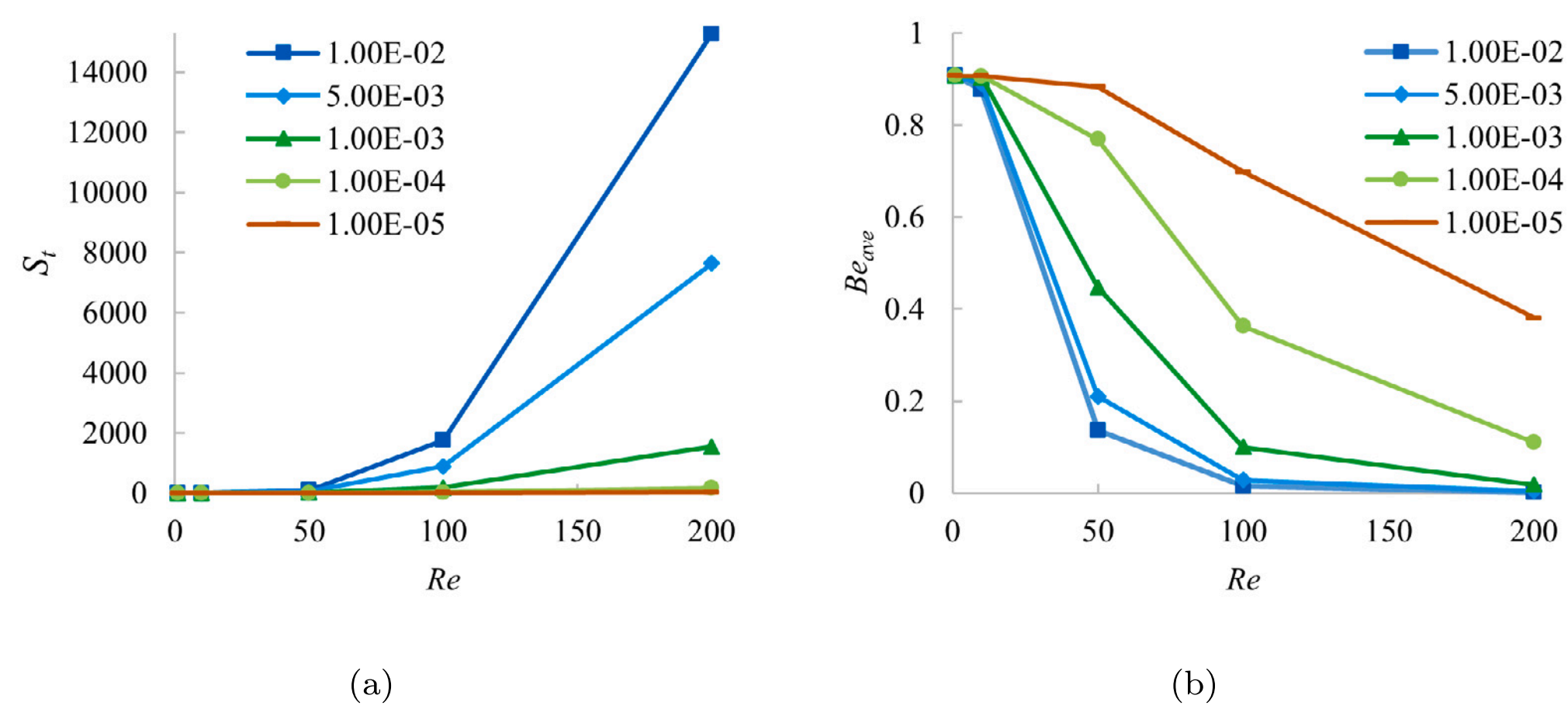

**Fig. 15.** Effects of irreversibility factor, Θ, for **Case 2** on (a) average entropy generation, $S_t$; (b) average Bejan number, $Be_{ave}$.

Fig. 18 illustrates the impact of wavy wall peaks ($n$) on **Case 3**. The presence of wavy walls introduces additional complexity, with streamlines showing more pronounced vortex structures at higher $n$, indicating enhanced mixing and flow disruption. Isotherms demonstrate more asymmetric thermal distributions due to the wavy walls, particularly at higher $n$, leading to more efficient heat transfer in localized regions. Entropy generation is highest around the deflector and the cylinders, with $S_l$ becoming more dispersed as $n$ increases. The Bejan number contours highlight a nuanced balance between heat transfer and fluid friction, with $n$ enhancing the dominance of entropy generation driven by fluid friction, particularly around the deflector. These findings emphasize the significant impact of wavy walls in optimizing thermal systems by modifying flow and thermal fields, thereby enhancing fluid mixing and heat transfer efficiency.

Fig. 19 presents the local entropy generation contours for **Case 3**, displaying more complex patterns with higher entropy generation spread across the domain, reflecting more intricate interactions. The Bejan number highlights significant changes in the dominance of heat transfer and fluid friction, suggesting distinct flow and thermal conditions compared to the previous cases.

Fig. 20 explores the average Nusselt number variations for **Case 3**, incorporating additional parameters such as rotational speed ($\omega$) and incli-nation angle ($\alpha$), alongside nanoparticle volume fraction ($\phi$), Hartmann angle ($\gamma$), and wavy wall peaks ($n$). Fig. 20(a) reaffirms the positive impact of nanoparticle volume fraction on thermal performance, consistent across various flow conditions. Fig. 20(b) shows that increasing the Hartmann number enhances heat transfer, with a more significant effect observed at intermediate Hartmann angles, balancing magnetic damping and convective forces. Fig. 20(c) demonstrates that wavy wall peaks continue to enhance heat transfer by increasing fluid mixing. Fig. 20(d) and Fig. 20(e) introduce new insights: higher rotational speeds enhance mixing and heat transfer, while increasing inclination angles initially improve and then stabilize the Nusselt number. Compared to **Cases 1** and **2**, the addition of the flow deflector and rotational effects in **Case 3** leads to the most complex flow and thermal interactions, resulting in the highest Nusselt numbers overall. These findings provide new insights into the multifaceted influences of geometric and dynamic factors in optimizing the thermal management of nanofluid systems, demonstrating how each case builds upon the previous in complexity and thermal performance enhancement.

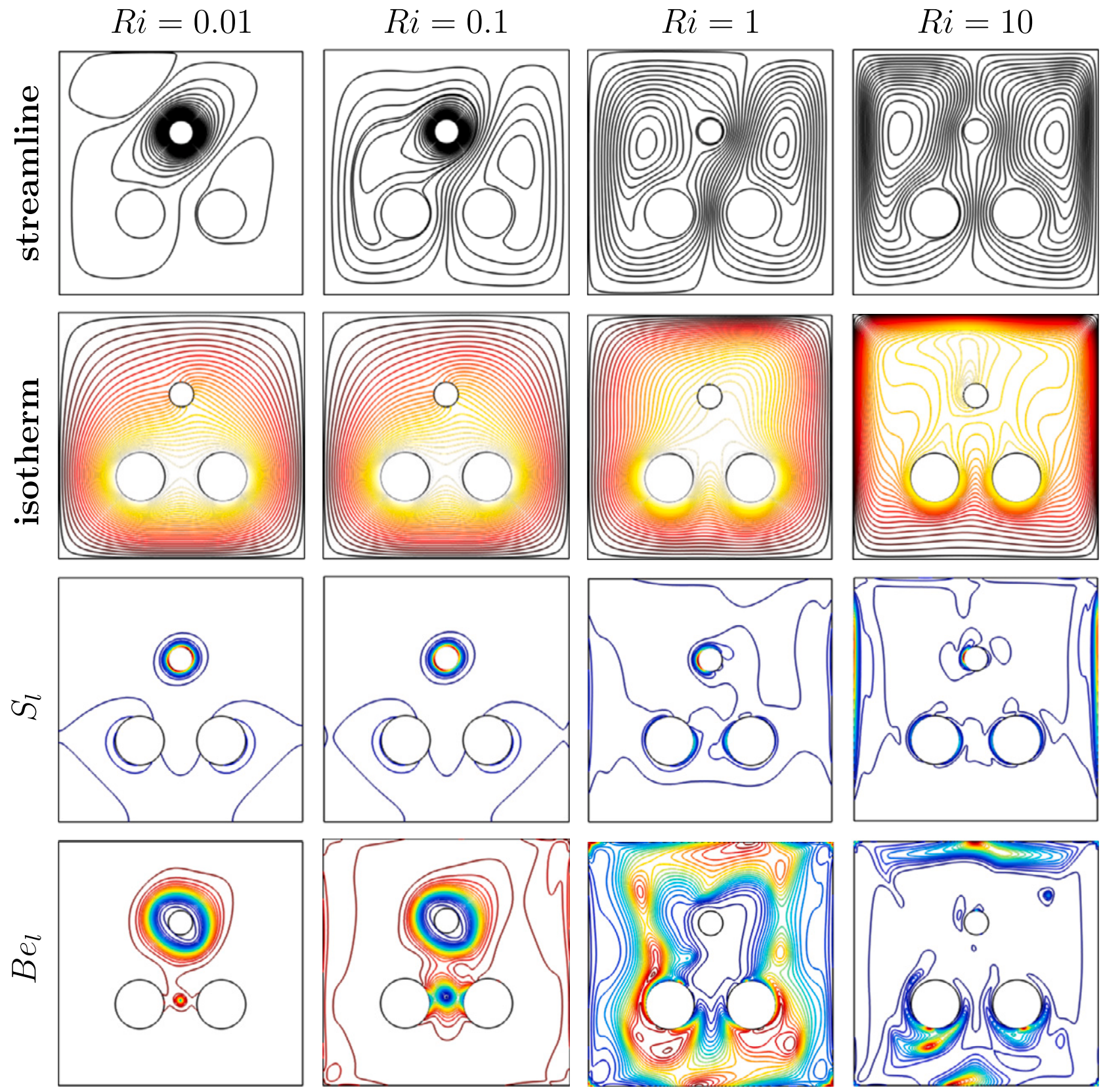

**Fig. 16.** Effects of Richardson number ($Ri$) variations on streamlines, isotherms, local entropy generation, and local Bejan number when $Re = 100$, $Ha = 20$, $\phi = 0.06$, $\omega = 1$, and $\gamma = 30°$ for **Case 3**.

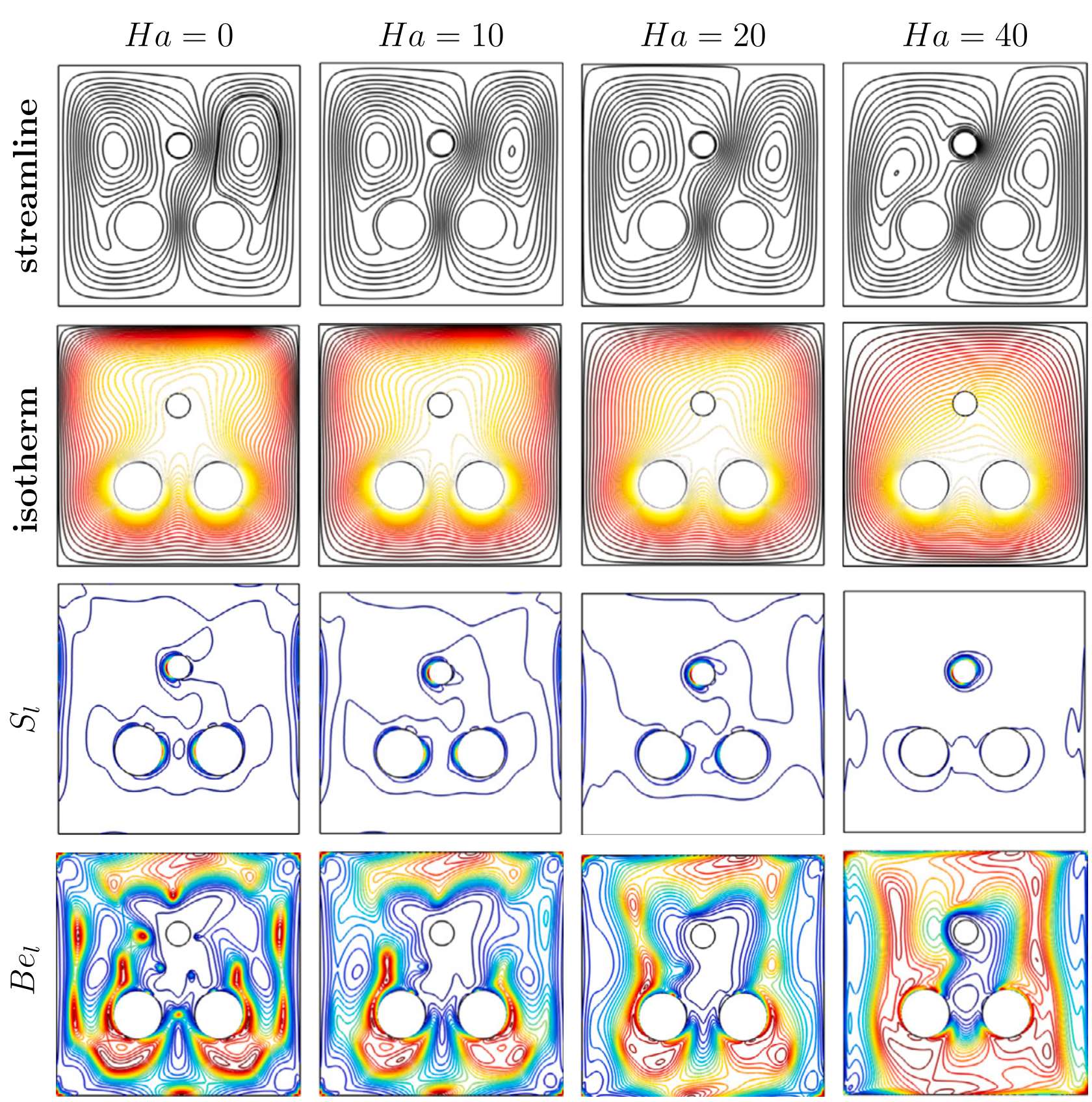

**Fig. 17.** Effects of Hartmann number ($Ha$) variations on streamlines, isotherms, local entropy generation, and local Bejan number when $Re = 100$, $Ri = 1$, $\phi = 0.06$, $\omega = 1$, and $\gamma = 30°$ for **Case 3**.

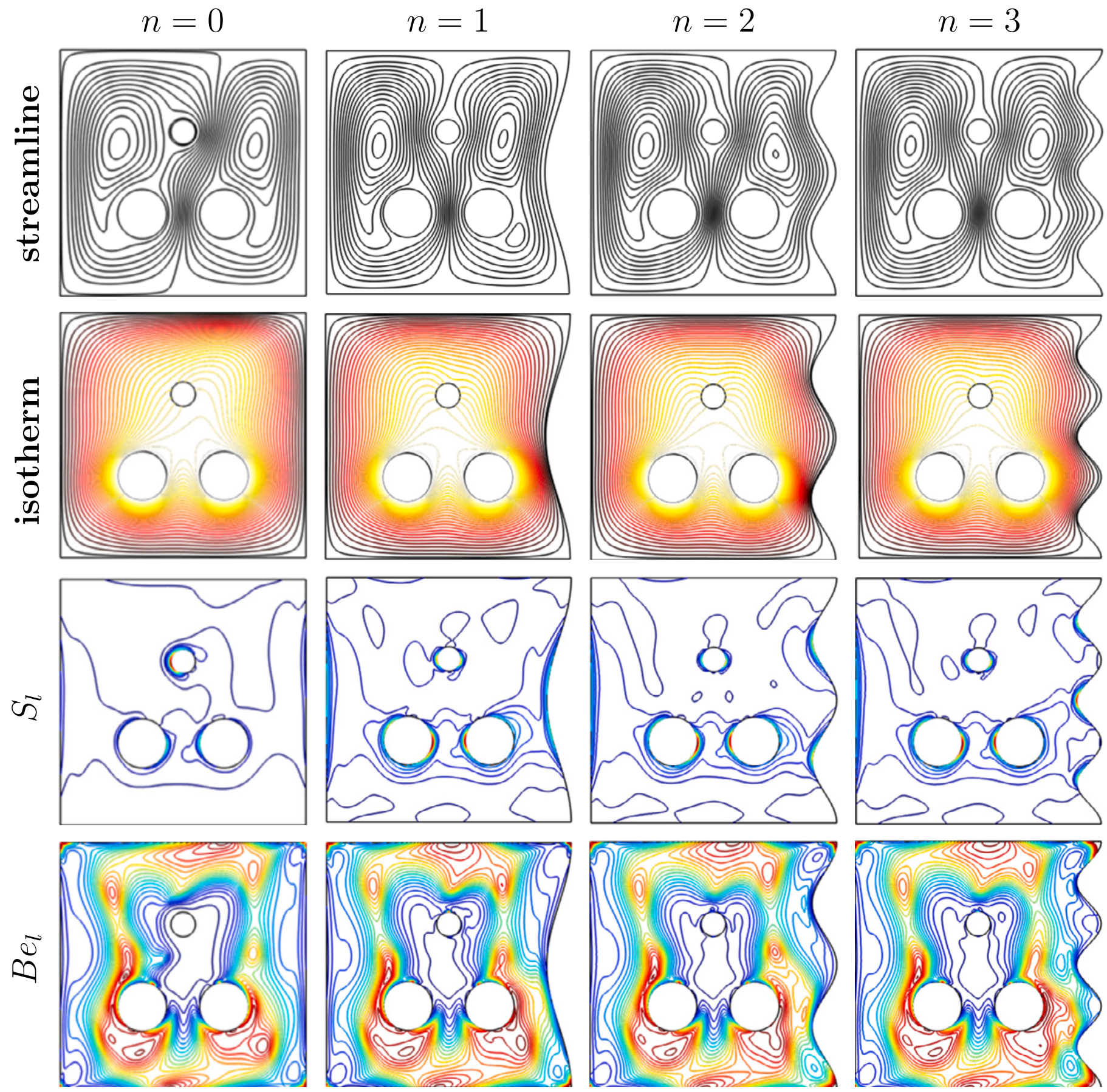

**Fig. 18.** Effects of wavy wall peaks ($n$) variations on streamlines, isotherms, local entropy generation, and local Bejan number when $Re = 100$, $Ri = 1$, $Ha = 20$, $\phi = 0.06$, $\omega = 1$, $a = 0.05$, and $\gamma = 30°$ for **Case 3**.

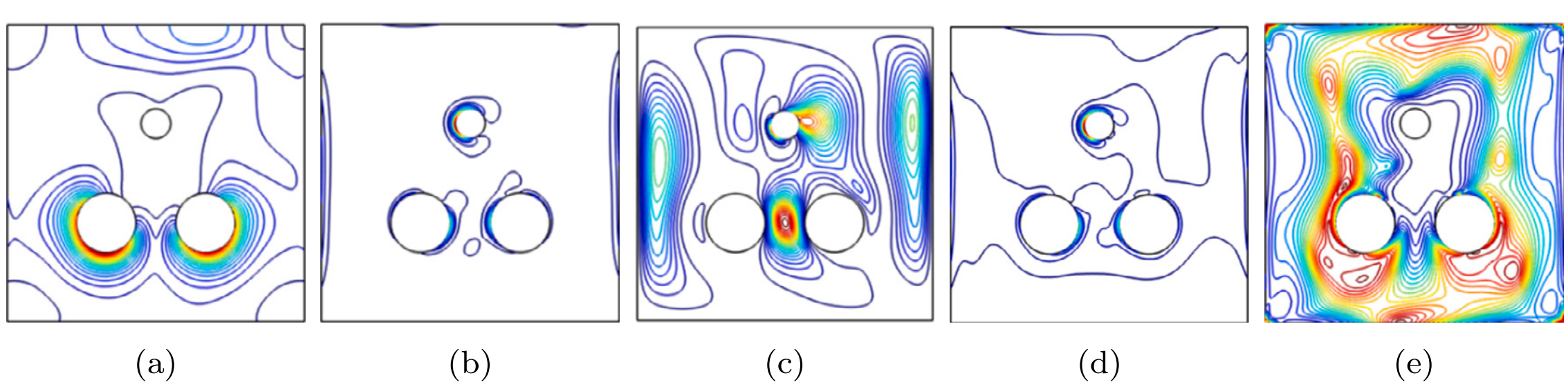

**Fig. 19.** Local entropy generation contours when $Re = 100$, $Ri = 1$, $Ha = 20$, $\phi = 0.06$, $\omega = 1$, and $\gamma = 30°$ for **Case 3**; (a) local entropy generation due to heat transfer, (b) local entropy generation due to fluid friction, (c) local entropy generation due to Joule heating, (d) local entropy generation, (e) local Bejan number.

Fig. 21 reveals that entropy generation ($S_t$) is higher and more complex due to the presence of a flow deflector in **Case 3**, with sharp increases at higher $Re$ and $\Theta$. Comparing **Case 3** with **Case 2**, the flow deflector results in a more intricate interaction, leading to increased $S_t$ and a more pronounced decrease in the average Bejan number ($Be_{\text{ave}}$). This indicates that the flow deflector enhances the contribution of fluid friction to overall entropy generation, surpassing that of heat transfer.

Table 5 reveals distinct performance differences between SWCNT and MWCNT nanofluids across the three cases. MWCNT consistently exhibits higher Nusselt numbers, indicating superior heat transfer capabilities with higher entropy generation values, indicating greater irreversibility compared to SWCNT, which exhibits higher Bejan num-bers signifying a greater dominance of thermal entropy generation over fluid friction. The higher entropy generation in SWCNT nanofluids is attributed to their lower thermal conductivity and higher thermal resistance compared to MWCNTs. This trend is most pronounced in Case 2, where the introduction of a second cylinder enhances thermal interactions by increasing velocity gradients and promoting stronger vortex formation. The flow deflector in Case 3 further influences heat transfer by redirecting flow, reducing stagnant regions, and improving temperature uniformity. As reflected in Table 5, $Nu_{ave}$ increases from 6.33 (Case 1) to 10.69 (Case 2) for MWCNT nanofluid, demonstrating the effect of the second cylinder on convective enhancement. The addition of a flow deflector in Case 3 results in $Nu_{ave} = 10.02$, still significantly higher than Case 1 but slightly lower than Case 2, indicating a shift in heat transfer mechanisms due to flow redirection. Similarly, entropy generation ($S_t$) follows an increasing trend, highlighting the trade-off between improved heat transfer and increased irreversibility. These findings align with previous studies [1,55], confirming that nanofluid selection and system geometry significantly influence heat transfer performance. While MWCNT nanofluids offer superior convective heat transfer, they also lead to increased entropy generation, necessitating a trade-off between efficiency and irreversibility. This insight is particularly valuable for applications such as microelectronics cooling and energy storage, where precise thermal control is essential.

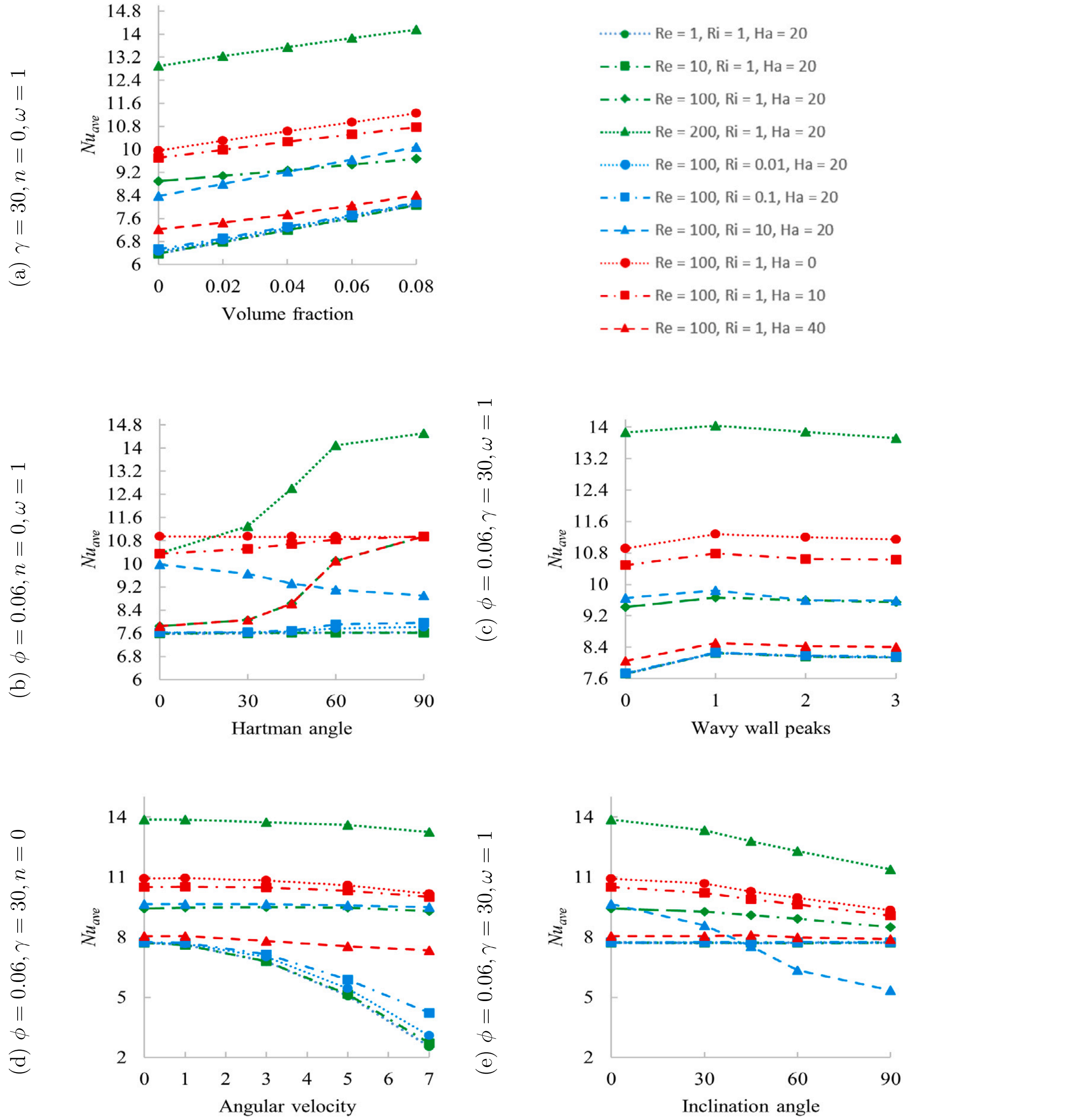

**Fig. 20.** Average Nusselt number variations for **Case 3** due to effects of (a) volume fraction of nanoparticles, $\phi$; (b) Hartmann angle, $\gamma$; (c) wavy wall peaks, $n$; (d) rotational speed, $\omega$; and (e) inclination angle, $\alpha$.

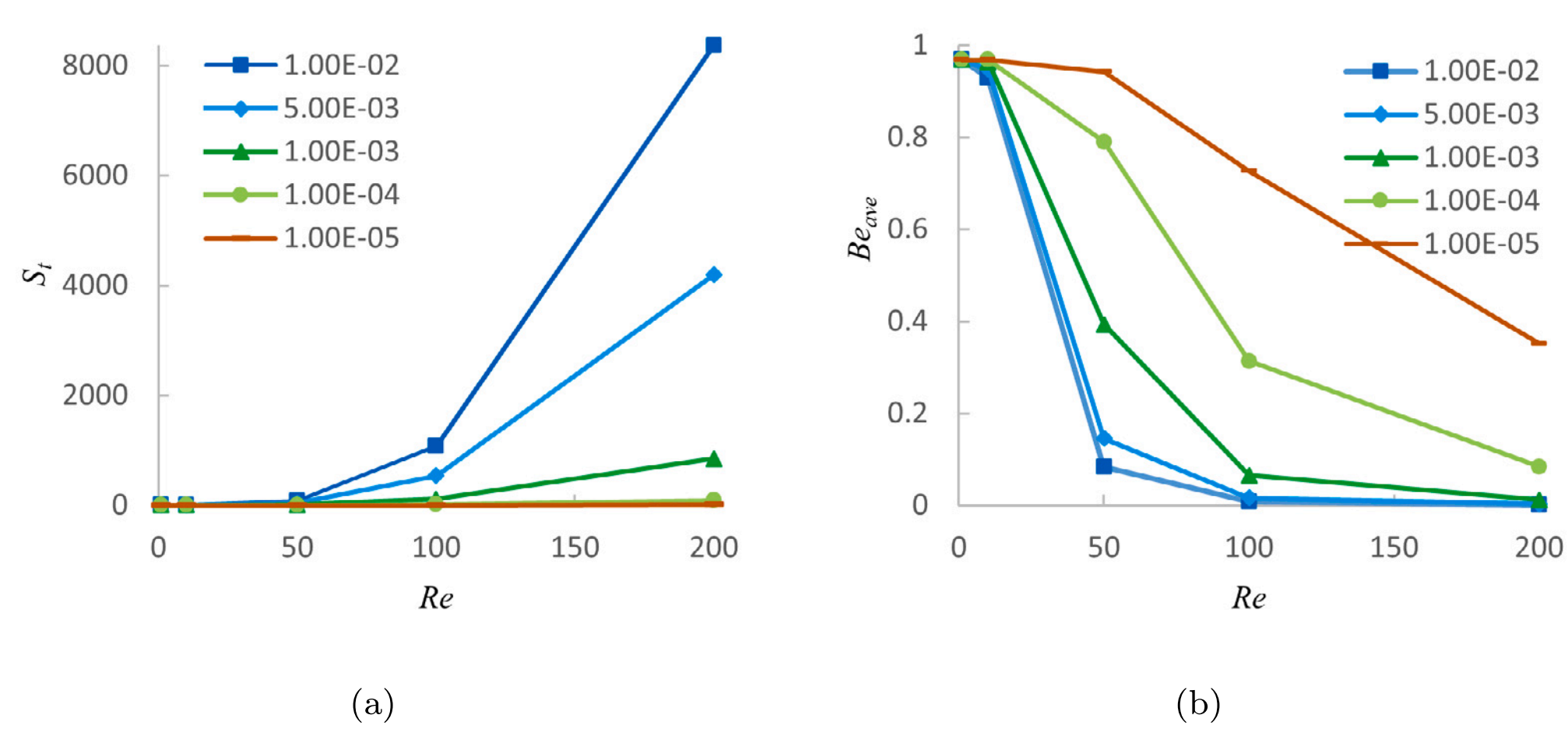

**Fig. 21.** Effects of irreversibility factor, $\Theta$, for **Case 3** on (a) average entropy generation, $S_t$; (b) average Bejan number, $Be_{ave}$.

**Table 5**
Estimated values based on fluid characteristics when $Re = 100$, $Ri = 1$, $Ha = 20$, $\phi = 0.06$, $\omega = 1$, and $\gamma = 30$.

| Parameters | Case 1 | | Case 2 | | Case 3 | |
|---|---|---|---|---|---|---|
| | SWCNT | MWCNT | SWCNT | MWCNT | SWCNT | MWCNT |
| $Nu_l$ | 5.5742 | 6.3342 | 9.5699 | 10.691 | 9.0289 | 10.026 |
| $Nu_{ave}$ | 5.5719 | 6.3316 | 9.5661 | 10.686 | 9.0252 | 10.022 |
| $S_l$ | 16.530 | 22.219 | 28.486 | 39.192 | 29.268 | 39.199 |
| $S_t$ | 16.531 | 22.220 | 27.565 | 37.925 | 28.086 | 37.616 |
| $Be_l$ | 0.3132 | 0.2387 | 0.3753 | 0.3202 | 0.3845 | 0.3323 |
| $Be_{ave}$ | 0.3132 | 0.2387 | 0.3632 | 0.3116 | 0.3690 | 0.3189 |

Several experimental studies support the trends observed in our numerical simulations. Abreu et al. [60] reported a 47% enhancement in convective heat transfer using MWCNT-water nanofluids under laminar flow conditions. Palanisamy and Kumar [61] observed up to 68% higher Nusselt numbers in a cone helically coiled tube heat exchanger with MWCNT nanofluids, emphasizing the impact of nanoparticle concentration and flow geometry. Lyu et al. [62] demonstrated enhanced thermal and fluid dynamic performance of MWCNT-water nanofluids in microchannel heat sinks with negligible increase in pumping power. Tetik et al. [63] found a 28% rise in thermal conductance using MWCNT–$SiO_2$ hybrid nanofluids in radiator systems, and Oliveira et al. [64] highlighted the sensitivity of heat transfer to nanoparticle loading and operating conditions. Building upon these experimentally supported trends, our work provides further insights by quantifying thermal irreversibility through Bejan number analysis, which reveals that SWCNT nanofluids exhibit substantially higher Bejan numbers than MWCNT nanofluids, indicating a greater dominance of thermal entropy generation over fluid friction. Additionally, we examine the influence of magnetic field effects, showing that increasing the Hartmann number ($Ha$) stabilizes the flow and enhances convective heat transfer, especially in geometrically modified configurations. These findings complement existing experimental literature and extend the understanding of transport characteristics in nanofluid-based thermal systems.

## 5. Conclusions

This study comprehensively investigates the mixed convection of CNT-water nanofluid in a cavity with heated cylinders under the influence of a magnetic field, focusing on three different geometric configurations: a single heated cylinder, two heated cylinders, and two heated cylinders with a flow deflector. The analysis encompasses the effects of the Reynolds number (Re), Richardson number (Ri), Hartmann number (Ha), wavy wall peaks (n), volume fraction of nanoparticles ($\phi$), Hartmann angle ($\gamma$), rotational speed ($\omega$), and inclination angle ($\alpha$) on the thermal and fluid dynamic behavior of the system.

MWCNT nanofluids consistently exhibit higher Nusselt numbers than SWCNT nanofluids, with up to a 19.1% increase in $Nu_{ave}$, confirming their superior heat transfer performance. Adding a second heated cylinder increases $Nu_{ave}$ by 71.7%, enhancing thermal interactions through stronger velocity gradients and vortex formation. MWCNT nanofluids exhibit higher entropy generation due to their superior thermal conductivity, while SWCNT nanofluids exhibit Bejan numbers up to 58.7% higher, indicating greater thermal irreversibility. The flow deflector further improves temperature uniformity by modifying vortex structures and reducing stagnant regions, optimizing convective transport. These findings emphasize the role of nanofluid selection and geometric modifications in balancing heat transfer efficiency and entropy generation.

The interaction between nanoparticles and magnetic fields provides a means to control heat transfer and flow stability. Increasing the Hartmann number stabilizes the flow as the Lorentz force suppresses turbulence and dampens velocity fluctuations, leading to a more uniform thermal field. The Hartmann angle ($\gamma$) also influences heat transfer efficiency, with intermediate angles providing optimal performance by aligning the magnetic field to balance flow suppression and convective transport. The introduction of wavy wall peaks enhances fluid mixing and heat transfer, particularly at higher Reynolds numbers and Hartmann numbers due to stronger secondary flow interactions and thermal boundary layer disruption. Geometric modifications such as additional cylinders and flow deflectors significantly influence the balance between heat transfer and fluid friction entropy generation. The flow deflector in Case 3 notably increases the complexity of vortex structures, which enhances localized convective heat transport by directing flow toward high-temperature gradients.

Increasing wavy wall peaks ($n$) enhances convective heat transfer by intensifying vortex formation, disrupting thermal boundary layers, and promoting fluid mixing, leading to improved thermal performance. Higher Reynolds numbers further amplify these effects, resulting in the most significant improvements in Nusselt number. Rotational speed ($\omega$) and inclination angle ($\alpha$) also influence heat transfer, with higher rotational speeds enhancing mixing and specific inclination angles minimizing flow separation, ensuring a stable convective transport mechanism and enhancing thermal performance by optimizing the interaction between thermal and inertial forces.

This study provides a comprehensive assessment of the interplay between nanofluid properties, geometric modifications, and magnetic field effects on mixed convection and entropy generation. While MWCNT nanofluids demonstrate superior heat transfer, they also contribute to higher entropy generation, requiring a trade-off between efficiency and irreversibility. The irreversibility factor ($\Theta$) has been analyzed to quantify this balance, offering new insights into the energy dissipation characteristics of different configurations. Furthermore, our findings align with prior studies [1,55], confirming the role of magnetic fields in stabilizing flow structures and enhancing heat transfer efficiency. These insights are particularly valuable for designing next-generation thermal management systems, including microelectronics cooling, energy storage, and aerospace applications.

Future studies could explore hybrid nanofluids, leveraging multiple nanoparticles for enhanced thermophysical properties and heat transfer. Artificial intelligence models, such as ANNs and CNNs, could provide predictive frameworks for optimizing heat transfer and entropy generation, enabling real-time parameter selection in industrial and aerospace applications. Further investigation of dynamic boundary conditions, vibration effects, and complex geometries could offer deeper insights into controlling thermal efficiency.

## CRediT authorship contribution statement

**Hashnayne Ahmed:** Writing – review & editing, Writing – original draft, Visualization, Validation, Software, Resources, Methodology, Investigation, Formal analysis, Data curation, Conceptualization. **Shashanka Biswas:** Writing – review & editing, Writing – original draft, Investigation, Formal analysis, Conceptualization. **Farzana Akter Tina:** Writing – review & editing, Writing – original draft.

## Funding

Self-funded.

## Declaration of competing interest

The authors declare that they have no known competing financial interests or personal relationships that could have appeared to influence the work reported in this paper.

## Data availability

Data will be made available on request.